\RequirePackage{ifpdf}
\ifpdf % We are running pdfTeX in pdf mode
\documentclass[pdftex]{sigma}
\else
\documentclass{sigma}
\fi

\numberwithin{equation}{section}

\begin{document}

\allowdisplaybreaks

\renewcommand{\thefootnote}{$\star$}

\renewcommand{\PaperNumber}{072}

\FirstPageHeading

\ShortArticleName{Appell Transformation and Canonical Transforms}

\ArticleName{Appell Transformation and Canonical Transforms\footnote{This paper is a
contribution to the Special Issue ``Symmetry, Separation, Super-integrability and Special Functions~(S$^4$)''. The
full collection is available at
\href{http://www.emis.de/journals/SIGMA/S4.html}{http://www.emis.de/journals/SIGMA/S4.html}}}

\Author{Amalia TORRE}

\AuthorNameForHeading{A.~Torre}

\Address{ENEA UTAPRAD-MAT Laboratorio di Modellistica Matematica, \\
via E. Fermi 45, 00044 Frascati (Rome), Italy}
\Email{\href{mailto:amalia.torre@enea.it}{amalia.torre@enea.it}}

\ArticleDates{Received January 31, 2011, in f\/inal form July 11, 2011;  Published online July 19, 2011}

\Abstract{The interpretation of the \textit{optical} Appell
transformation, as previously elaborated in relation to the free-space
paraxial propagation under both a rectangular and a circular cylindrical
symmetry, is reviewed. Then, the \textit{caloric} Appell transformation,
well known in the theory of heat equation, is shown to be amenable for a
similar interpretation involving the Laplace transform rather than the
Fourier transform, when dealing with the 1D heat equation. Accordingly, when
considering the radial heat equation, suitably def\/ined Hankel-type
transforms come to be involved in the inherent Appell transformation. The
analysis is aimed at outlining the link between the Appell transformation
and the canonical transforms.}


\Classification{35K05; 35K10; 47D06}

\renewcommand{\thefootnote}{\arabic{footnote}}
\setcounter{footnote}{0}

\section{Introduction}

The Appell transformation \cite{appell} is mainly associated with the heat
equation (HE) \cite{appell,widderb,rosenbloom,widder1,widder2,leutwiler,shimomura}.
Some recent analyses however
have shown its relevance in other contexts as well \cite{brzezina,torreat}.

In \cite{torreat} some issues related to the 1D~HE have been revisited
within the context of the free-space paraxial propagation, formally
accounted for by the 2D paraxial wave equation (PWE). Thus, the Appell
transformation has been interpreted in the light of the propagation of given
source functions, which are in a def\/inite relation with the source functions
of the original \textit{wavefunctions} (i.e.\ solutions of the PWE),
relatively to both a rectangular and a circular cylindrical symmetry.

The analysis has been developed by following the Lie-algebra based approach
to evolution equations, ruled by Hamiltonian operators underlying a harmonic
oscillator-like symmetry algebra, as originally elaborated in a series of
seminal papers by Kalnins, Miller and Boyer \cite{kalnins1,kalnins2}.

In fact, the 1D HE and the 2D PWE  as well as the (1+1)D Schr\"{o}dinger
equation (SE) can be considered as evolution equations, ruled by
Hamiltonian-like operators (not necessarily Hermitian) which are quadratic
in the inherent canonically conjugate variables.

The 1D~HE has been the object of deep investigations
 \cite{appell,widderb,rosenbloom,widder1,widder2,leutwiler,
shimomura,brzezina,torreat}, with the relevant Appell transformation being also embedded in
the Lie symmetry group \cite{miller,wolfb,olver}. Extensive studies have
been devoted to the PWE and the SE as well. In \cite{torrepwe} and~\cite{torresp}, for instance, the aforementioned Lie algebra based method has
been systematically applied to the 2D PWE, and markedly interpreted in terms
of solutions obtained by propagating def\/inite ``source functions'',
identif\/ied according to that method as eigenstates of specif\/ic operators in
the inherent symmetry algebra \cite{kalnins1,kalnins2,miller}. Consequently,
the transformations between wavefunctions can be seen to trace back to
def\/inite relations between the respective source functions, as explicitly
shown in~\cite{torrel} through an analysis aimed at characterizing
transformations between wavefunctions in terms of transformations between
the relevant source functions.

In particular, the \textit{optical} Appell
transformation represents the mapping between wavefunctions resulting from
Fourier or Hankel pairs of source functions. This has been shown in \cite{torreat} by proving the correspondence through the Appell transformation of
 wavefunctions generated by eigenstates of operators belonging to
the PWE symmetry algebra, which are linked by a Fourier or Hankel
similarity transformation, according to whether a rectangular or circular
cylindrical geometry is concerned. As a mere consequence, the fractional Appell
transformation has been introduced by involving the fractional Fourier or
Hankel transform. Thus, a family of symmetry transformations for the 2D
PWE, parameterized by a continuous parameter, has been identif\/ied.

Although a direct check of the result can
easily be carried out, the formal steps followed through the analysis, as
presented in~\cite{torreat} and reviewed below, have been aimed at favoring
a~``visualization'' of the process, in view also of the possibility of
suggesting a practical scheme for the realization of the optical ``Appell
transformer'' as parallel to the optical Fourier transformer.

The purpose of the present paper is to show that a similar interpretation of
the \textit{caloric} Appell transformation can be elaborated, resorting to
the bilateral Laplace transform rather than to the Fourier transform, as far
as the 1D HE is concerned. We will frame the analysis within the Lie-algebra
based formalism, as it~-- besides its intrinsic formal and conceptual
elegance~-- reveals ef\/fective in suggesting relations with other f\/ields or
theories. By retracing the same formal steps followed when dealing with the
PWE, we will prove in fact that the \textit{caloric} Appell transformation
manifests the correspondence between temperature functions generated by
eigenstates of operators belonging to the HE symmetry algebra, which are
linked by a Laplace-similarity transformation. Then, resorting to the
def\/inition of the fractional Laplace transform, a \textit{fractional caloric}
Appell transformation will as well be introduced.

Although, as earlier noted, the caloric Appell transformation has already
been embedded in the Lie symmetry group \cite{miller,wolfb,olver}, to the
author's knowledge, the Laplace-based relation of concern in the present
context has still not been explicitly discussed in the literature. In our
opinion, indeed, the parallelism between the HE and the PWE, which
stems from the underlying algebra, is still not thoroughly exploited, whilst
it may yield interesting results. For instance, the property, proven in~\cite{leutwiler}, that the Appell transformation for an in general $n$-dimensional HE  is \textit{essentially} the only
symmetry transformation (in the sense that every symmetry transformation can
be obtained by composing Appell transformations with suitable scalings and
shifts of all the variables) has been extended in~\cite{torren} to the 2D PWE as well, by resorting to the
symplectic ray-matrix formalism, usually adopted in paraxial optics. The
stated property stems from the fact that the generators of the PWE symmetry
algebra are Fourier or Hankel-similarity related.

Likewise, in regard to the radial HE we will prove that the relevant Appell
transformation involves a similarity transformation by properly def\/ined
Hankel-type transforms.

As a result, one can recognize a direct link between the Appell
transformation and the linear canonical transforms, since the (fractional)
Fourier transform is an important representative of the real linear
canonical transforms as the (fractional) bilateral Laplace transform is an
equally important representative of the complex linear canonical transforms.
Likewise, as we will see, the transforms, which relate to the Appell
transformation for the radial HE, can be framed within the context of the
radial canonical transforms.

In Section~\ref{section2} we will f\/irstly review the basics of the symmetry algebra-based
method and later the results presented in \cite{torreat}, where the \textit{optical} Appell transformation has been introduced and related to the
Fourier or Hankel transform according to whether the PWE in rectangular or
radial coordinate is concerned. In Section~\ref{section3} the basics of the canonical
transforms theory, specif\/ically in relation with the bilateral Laplace and
Hankel-type transforms, will be reviewed. Then, retracing the steps of the
analysis presented in Section~\ref{section2}, we will prove in Section~\ref{section4} that the \textit{caloric} Appell transformation for the 1D HE connects temperature functions
generated by eigenstates of operators in the inherent symmetry-algebra,
which are linked through a Laplace-similarity transformation. A similar
relation will be shown to hold also for the Appell transformation for the
radial HE, accordingly resorting to suitably def\/ined Hankel-type transforms,
explicable as ``radial-Laplace''-type transforms. Concluding comments will
be given in Section~\ref{section5}.

\section{Optical Appell transformation:\\  Fourier and Hankel transforms}\label{section2}

There is a well-known formal analogy between the (2+1)D SE, ruling the
dynamics in two space-dimensions of a quantum particle under the action of
some potential, and the 3D PWE, describing the paraxial propagation of a
monochromatic scalar light-f\/ield through some medium, under the basic
correspondences $t\rightarrow z$, $\hbar \rightarrow \lambda /2\pi $, and
some others less direct, relating the momentum of the particle and the
potential to the direction of propagation of the signal and the refractive
index of the medium. Due to such an analogy, in fact, the Lie algebra-based
method, as developed in a series of seminal papers by Kalnins, Miller and
Boyer~\cite{kalnins1,kalnins2,miller} in connection with the solutions of the
SE, has been applied as well to the free-space 2D PWE, both ``linear'' and
``radial''~\cite{torrepwe,torresp}, on account in particular of the results
in~\cite{kalnins1}, pertaining the (1+1)D SE respectively for a free
particle and for a particle under a potential of the type~$c/x^2$.

The free-space paraxial propagation is accounted for by the (free-space) 3D
PWE, conveniently expressed in the normalized form:
\begin{gather}
\left[ 2i\frac \partial {\partial \zeta }+\frac{\partial ^2}{\partial \xi ^2%
}+\frac{\partial ^2}{\partial \eta ^2}\right]v(\xi ,\eta ,\zeta )=0,
\label{normparg}
\end{gather}
the \textit{wavefunction} $v(\xi ,\eta ,\zeta )$ being intended to belong to
the complex space $\mathfrak{F}$ of locally $C^\infty $ functions of the real
unitless variables~$\xi $, $\eta $, $\zeta $. The latter represent the
transverse and longitudinal (Cartesian) coordinates~$x$,~$y$, and~$z$
scaled, respectively, to some characteristic transverse scale~$w_0$, and to
the relevant confocal parameter $b=kw_0^2$, $k$ being the wavenumber of the
light.

Equation (\ref{normparg}) admits separable-variable solutions in both
Cartesian and polar coordinates, which are suitable to describe
wavefunctions displaying respectively a rectangular and a circular
cylindrical symmetry. In both cases one deals with a 2D equation, involving
only one transverse coordinate, the longitudinal coordinate $\zeta $ playing
the role of an evolution variable.

\subsection{2D ``linear''  paraxial wave equation}\label{section2.1}

The separability of the solutions of (\ref{normparg}) in rectangular
coordinates manifests in the factorization of the 3D wavefunction $v(\xi
,\eta ,\zeta )$ in terms of two 2D wavefunctions $u(\xi ,\zeta )$ and $%
w(\eta ,\zeta )$ as
\[
v(\xi ,\eta ,\zeta )=u(\xi ,\zeta )w(\eta ,\zeta ),
\]
each satisfying the 2D PWE in one transverse coordinate:
\begin{gather}
\left[ 2i\frac \partial {\partial \zeta }+\frac{\partial ^2}{\partial \xi ^2}\right]u(\xi ,\zeta )=0.  \label{normpar}
\end{gather}
It is formally similar to the (1+1)D SE for the free-particle Hamiltonian $\widehat{\mathrm{H}}=-(1/2)\partial ^2/\partial \xi ^2$.

The relevant evolution operator (i.e.\ the \textit{paraxial} \textit{propagator}) is
\begin{gather*}
\widehat{\mathcal{U}}_{_{\mathsf{PWE}}}(\zeta ):=e^{i\frac \zeta 2 \frac{\partial ^2}{\partial \xi ^2}},  %\label{propagator}
\end{gather*}
so that solutions to (\ref{normpar}), for given initial conditions $u(\xi
,0)=u_0(\xi )$, are obtained as \cite{kato}{\samepage
\begin{gather}
u(\xi ,\zeta )=\big[\widehat{\mathcal{U}}_{_{\mathsf{PWE}}}(\zeta )u_0\big](\xi
)=\frac 1{\sqrt{2\pi i\zeta }}\int_{-\infty }^\infty e^{i\frac{(\xi -\xi
^{\prime })^2}{2\zeta }}u_0(\xi ^{\prime })d\xi ^{\prime },  \label{hfi}
\end{gather}
under the minimal assumption that $u_0(\xi )$ tends to zero suf\/f\/iciently
rapidly as $\xi \rightarrow \pm \infty $.}

The above integral is well-known in paraxial optics as the Huygens--Fresnel
(or, Collins) dif\/fraction integral (specif\/ically, for
free-propagation)~\cite{collins,siegman}.

 Equation (\ref{hfi}) implicitly assumes the equivalence of the
representations of the propaga\-tor~$\widehat{\mathcal{U}}_{_{\mathsf{PWE}}}(\zeta )$ as a Fresnel transform and as an exponential operator, involving
the free-Hamilto\-nian. Evidently, the former is meaningful only if $u_0(\xi )$
is integrable and the pertinent integral converges, whereas the latter
requires the implied series of derivatives~$\frac{\partial ^{2n}u_0}{\partial \xi ^{2n}}$ to exist and converge to a f\/inite value as well. We
will not dwell here on the legitimacy of such an equivalence, but we will
simply assume to deal with functions for which it holds~\cite{kato}.

\subsubsection[Solving the PWE: the Kalnins-Miller-Boyer method]{Solving the PWE: the Kalnins--Miller--Boyer method}\label{section2.1.1}

The symmetry-algebra based approach, as developed in~\cite{kalnins1,kalnins2,miller},
identif\/ies a systematic method to solve (\ref{normpar}), based on
the spectral decomposition of operators belonging to the inherent symmetry
algebra $\mathcal{G}$. This is the semidirect sum of the algebra $sl(2,\mathbb{R})\simeq sp(2,\mathbb{R})\simeq su(1,1)$, spanned by the operators $\big\{
\widehat{\mathrm{K}}_{+},\widehat{\mathrm{K}}_3, \widehat{\mathrm{K}}%
_{-}\big\} $, with
\begin{gather}
\widehat{\mathrm{K}}_{+}:=\frac 12\xi ^2,\qquad \widehat{\mathrm{K}}_3:=-\frac i2\left(\xi \frac \partial {\partial \xi }+\frac 12\right) ,\qquad \widehat{\mathrm{K}}_{-}:=-\frac 12\frac{\partial ^2}{\partial \xi ^2},\nonumber\\
\big[ \widehat{\mathrm{K}}_{+}, \widehat{\mathrm{K}}_{-}\big]  =2i \widehat{\mathrm{K}}_3,\qquad \big[ \widehat{\mathrm{K}}_{\pm },\widehat{\mathrm{K}}_3\big]  =\pm  i\widehat{\mathrm{K}}_{\pm },\label{su11}
\end{gather}
and the Weyl algebra $\mathcal{W}$, spanned by $\big\{\widehat{\mathrm{X}},\widehat{\mathrm{P}},\widehat{\mathrm{I}}\big\}$, such that
\begin{gather}
\widehat{\mathrm{X}}:=\xi ,\qquad \widehat{\mathrm{P}}:=-i\frac \partial
{\partial \xi },\qquad \widehat{\mathrm{I}}=1,\qquad \big[ \widehat{\mathrm{X}},\widehat{\mathrm{P}}\big] =i\widehat{\mathrm{I}}.  \label{weyl}
\end{gather}
The algebra $\mathcal{G}=\mathcal{W}\oplus sl(2,\mathbb{R})$ is sometimes referred to in
the literature as $wsl(2,\mathbb{R})$; $\mathcal{W}$ is an ideal in $wsl(2,\mathbb{R})$, as conveyed by the mixed commutators,
\begin{gather*}
\big[ \widehat{\mathrm{K}}_{+},\widehat{\mathrm{X}}\big]
 =0,\qquad \big[ \widehat{\mathrm{K}}_{+},\widehat{\mathrm{P}}\big]
 =i\widehat{\mathrm{X}}, \qquad \big[ \widehat{\mathrm{K}}_{-},\widehat{\mathrm{X}}\big]
 =-i\widehat{\mathrm{P}}, \qquad \big[ \widehat{\mathrm{K}}_{-},\widehat{\mathrm{P}}\big] =0,\nonumber \\
\big[ \widehat{\mathrm{K}}_3,\widehat{\mathrm{X}}\big]
=-\frac i2\widehat{\mathrm{X}},\qquad \big[ \widehat{\mathrm{K}}_3,\widehat{\mathrm{P}}\big]=\frac i2 \widehat{\mathrm{P}},
\qquad \big[ \widehat{\mathrm{K}}_{\pm ,3},\widehat{\mathrm{I}}\big] =0.%\label{wkcom}
\end{gather*}

The operators (\ref{su11}) and (\ref{weyl}) are self-adjoint when acting on
the Hilbert space $\mathfrak{L}^2(\mathbb{R})$ of complex-valued Lebesgue
square-integrable functions on the real line $\mathbb{R}$ with the scalar
product $\left\langle f,g\right\rangle \equiv \int_{-\infty }^{+\infty
}f(\xi )g^{*}(\xi )d\xi $, the star denoting complex conjugation.

As formulated in~\cite{kalnins1}, the spectral decomposition $\left\{
f_\lambda \right\} _\lambda $ of any $\widehat{\mathrm{K}}\in \mathcal{G}$,
\begin{gather}
\widehat{\mathrm{K}}f_\lambda (\xi )=\lambda f_\lambda (\xi ),  \label{bas}
\end{gather}
can be used to construct solutions of the PWE having the~$f_\lambda $s as
initial conditions.

This can be done by evaluating the Huygens--Fresnel integral (\ref{hfi}) for
each $f_\lambda $, provided it be in the domain of $\widehat{\mathrm{K}}_{-}$, thus obtaining the propagated functions $v_\lambda (\xi ,\zeta )=\widehat{\mathcal{U}}_{_{\mathsf{PWE}}}(\zeta )f_\lambda (\xi )$.

Alternatively, one can search for the solutions of the equations
\begin{gather}
\widehat{\mathrm{K}}(\zeta )v_\lambda (\xi ,\zeta )=\lambda v_\lambda (\xi
,\zeta )  \label{alt}
\end{gather}
which directly follow from (\ref{bas}) on account of $f_\lambda (\xi )=%
\widehat{\mathcal{U}}_{_{\mathsf{PWE}}}(\zeta )^{-1}v_\lambda (\xi ,\zeta )$%
, and hence address the initial conditions $v_\lambda (\xi ,0)=f_\lambda
(\xi )$. The operator $\widehat{\mathrm{K}}(\zeta )$, given by
\begin{gather}
\widehat{\mathrm{K}}(\zeta ):=e^{-i\zeta \widehat{\mathrm{K}}_{-}}\widehat{%
\mathrm{K}}e^{i\zeta \widehat{\mathrm{K}}_{-}},  \label{hop}
\end{gather}
can be understood as a Heisenberg-like operator corresponding to $\widehat{%
\mathrm{K}}$. In fact, the above mapping, which relates ``f\/ixed-location''
operators $\widehat{\mathrm{K}}\in \mathcal{G}$ to ``evolving-location''
operators $\widehat{\mathrm{K}}(\zeta )\in \mathcal{G}$, is reminiscent of
the transformation relating the Heisenberg and Schr\"{o}dinger pictures of
quantum mechanics (involving $\mathfrak{L}^2(\mathbb{R})$-wavefunctions)\footnote{The Heisenberg picture of quantum mechanics amounts to
the observable \cite{ballantine}
\[
\widehat{\mathrm{O}}(t)=\widehat{\mathcal{U}}^{\dagger }(t)\widehat{\mathrm{O%
}}\widehat{\mathcal{U}}(t),
\]
where $\widehat{\mathcal{U}}(t)$ is the unitary evolution operator, which
obeys the relations $\widehat{\mathcal{U}}^{\dagger }(t)=\widehat{\mathcal{U}%
}^{-1}(t)$ and $\widehat{\mathcal{U}}^{-1}(t)=\widehat{\mathcal{U}}(-t)$.}.
More precisely, the operator $\widehat{\mathrm{K}}(\zeta )$ should be
regarded as the ``back-evolving location'' form of $\widehat{\mathrm{K}}$.

Since
\[
e^{\widehat{\mathrm{A}}}\widehat{\mathrm{B}}e^{-\widehat{\mathrm{A}}}=\sum\limits_{j=0}^\infty ({\rm ad}\, \widehat{\mathrm{A}})^j\widehat{\mathrm{B}}/j!, \qquad \text{with} \quad ({\rm ad}\,\widehat{\mathrm{A}})^j\widehat{\mathrm{B}}=\big[ \widehat{\mathrm{A}}
,\big[\dots,\big[ \widehat{\mathrm{A}},\big[ \widehat{\mathrm{A}},\widehat{\mathrm{B}}\big] \big] \big] \big]
\] being a $j$-fold commutator,
the $\widehat{\mathrm{K}}(\zeta )$s are f\/inite linear combinations of the
f\/ixed-location ope\-ra\-tors (\ref{su11}) and (\ref{weyl}) with coef\/f\/icients
depending on the propagation variable $\zeta $, which so enters~(\ref{alt})
as a parameter.

Evidently, the above sketched procedure, based on the eigenvalue equations (\ref{bas}) or (\ref{alt}), is equivalent to a separation of variables; in
fact, in~\cite{kalnins1} a well-def\/ined connection between separation of
variables and Lie symmetries for (\ref{normpar}) has been established (see
also~\cite{miller}).

In \cite{torrepwe}, the procedure has markedly been interpreted in terms of
wavefunctions obtained by propagating def\/inite ``source functions'',
identif\/ied as eigenstates of specif\/ic operators in $\mathcal{G}$.
Accordingly, the transformations between wavefunctions trace back to
def\/inite relations between the respective source functions~\cite{torrel}.
Further cases besides those considered in~\cite{kalnins1} have been analyzed
in \cite{torrepwe}. Thus, a certain class of solutions of (\ref{normpar})
has been obtained by propagating the eigenfunctions of the operator $%
\widehat{\mathrm{K}}_{\xi _0}= \widehat{\mathrm{K}}_3-(2/\xi _0)\widehat{%
\mathrm{K}}_{+}$, $\xi _0$ being an arbitrary parameter. Such solutions
depend on three independent parameters and basically comprise a complex
quadratic exponential modulated by the Weber--Hermite function $D_{-2i\lambda
-1/2}$ of suitable argument.

The same result was previously deduced in \cite{bandresc} on the basis of an
ansatz giving the general solution of the 2D PWE in a suitable
separable-variable form. A similar ansatz have led the same authors to
identify general solutions of (\ref{normparg}) pertaining to a circular
cylindrical symmetry~\cite{bandresr}; the elliptical cylindrical symmetry
has been considered in \cite{bandrese}.

On the other hand, as shown in \cite{torrel}, a def\/inite relation between
the wavefunctions arising from the eigenfunctions of the operators $\widehat{%
\mathrm{K}}_{\xi _0}$ and $\widehat{\mathrm{K}}_3$ can be established, the
former following from the latter under the symmetry transformation produced
by the operator $\exp [(i/\xi _0)\widehat{\mathrm{K}}_{+}(\zeta )]$.

\subsubsection{Fourier transform and optical Appell transformation}\label{section2.1.2}

As proven in \cite{torreat}, the \textit{optical} Appell transformation
connects wavefunctions, whose source functions are Fourier related. In order
to review this result, we consider for every operator $\widehat{\mathrm{K}}%
\in \mathcal{G}$ the dual operator $\widehat{\widetilde{\mathrm{K}}}$,
linked to the former through a similarity transformation by the direct or
inverse Fourier transform operator $\widehat{\mathcal{F}}$, i.e.\ $\widehat{%
\widetilde{\mathrm{K}}}=\widehat{\mathcal{F}}\widehat{\mathrm{K}}\widehat{%
\mathcal{F}}^{-1}$ or $\widehat{\widetilde{\mathrm{K}}}=\widehat{\mathcal{F}}%
^{-1}\widehat{\mathrm{K}}\widehat{\mathcal{F}}$, both belonging to $\mathcal{%
G}$.

The Fourier transform, which is well known to operate as
\begin{gather*}
\widetilde{f}(x)=\big[\widehat{\mathcal{F}}f\big](x):=\frac 1{\sqrt{2\pi }
}\int_{-\infty }^{+\infty }e^{-ixx^{\prime}}f(x^{\prime })dx^{\prime },
%\label{fourierdef}
\end{gather*}
admits the exponential operator representation in terms of the $su(1,1)$
generators $\widehat{\mathrm{K}}_{+}$ and $\widehat{\mathrm{K}}_{-}$:
\begin{gather}
\widehat{\mathcal{F}}=e^{i\pi /4}e^{-i(\pi /2)[\widehat{\mathrm{K}}_{-}+\widehat{\mathrm{K}}_{+}]},  \label{fourier}
\end{gather}
the factor $e^{i\pi /4}$ allowing the optical transform to be matched to the
mathematical transform.

By optical transform we mean the linear canonical transform resulting from
the map of the symplectic group $Sp(2,\mathbb{R})$ ($\simeq SL(2,\mathbb{R})$)
into the metaplectic group $Mp(2,\mathbb{R})$, which turns optical ray-tracing
matrices $\mathbf{M}=\binom{A\,\,\,\,B}{C\,\,\,\,D}$ in the former into the
Collins integral in the latter as \cite{collins,siegman}
\begin{gather}
v(\xi ,\zeta _o)=\widehat{\mathcal{U}}_{\binom{A\,\,\,\,B}{C\,\,\,\,D}}v(\xi
,\zeta _i)=\frac 1{\sqrt{2\pi iB}}\int_{-\infty }^{+\infty }\exp \left\{\frac
i{2B}\big(A\xi ^{\prime 2}-2\xi \xi ^{\prime }+D\xi ^2\big)\right\}v(\xi ^{\prime
},\zeta _i)d\xi ^{\prime }. \!\!\!\! \label{uabcd}
\end{gather}
The optical system described by the ray-matrix $\mathbf{M}$ is therefore
seen as an operator transporting the wavefunction on the input plane located
at~$\zeta _i$ to the wavefunction on the output plane at~$\zeta _o$. The
metaplectic group $Mp(2,\mathbb{R})$ provides a double cover of the symplectic
group $Sp(2,\mathbb{R})$, to which the $\mathbf{M}$s belong. In fact, although
not explicitly displayed by convention, a double sign $\pm $ is implied in~(\ref{uabcd})\footnote{A double sign $\pm $ is accordingly implied also in
the matching factor entering~(\ref{fourier}).}, thus allowing the
metaplectic images of the~$\mathbf{M}$s to close into a group.

 It may be useful to clarify that throughout the paper $\sqrt{i}$ will be intended to
identify the principal square root of the imaginary unit: $\sqrt{i}=e^{i\pi /4}$, as it will be in general for the square root of any (complex) number: $\sqrt{a}=\sqrt{|a|}e^{i\arg (a)/2}$, with $\arg (a)\in (-\pi ,\pi]$.

The dif\/fraction integral (\ref{hfi}) follows from (\ref{uabcd}) in
correspondence to the free-section matrix
\begin{gather}
\mathbf{T}(\zeta )=
\begin{pmatrix}
1 & \zeta \\
0 & 1
\end{pmatrix}
,  \label{fsmatrix}
\end{gather}
i.e.\ $\widehat{\mathcal{U}}_{_{\mathsf{PWE}}}(\zeta )=\widehat{\mathcal{U}}_{\binom{1\,\,\,\,\zeta }{0\,\,\,\,1}}$.

In virtue of the disentanglement relation for the $su(1,1)$ algebra
generators
\begin{gather}
e^{i\beta [\widehat{\mathrm{K}}_{-}+\widehat{\mathrm{K}}_{+}]}=e^{i\tan
(\beta /2)\widehat{\mathrm{K}}_{-}}e^{i\sin (\beta )\widehat{\mathrm{K}}_{+}}e^{i\tan (\beta /2)\widehat{\mathrm{K}}_{-}}=e^{i\tan (\beta /2)\widehat{\mathrm{K}}_{+}}e^{i\sin (\beta )\widehat{\mathrm{K}}_{-}}e^{i\tan
(\beta /2)\widehat{\mathrm{K}}_{+}},  \label{dis}
\end{gather}
holding for $|\beta |<\pi $, the exponential operator in (\ref{fourier}) is
factorizable in the two equivalent forms
\begin{gather}
e^{-i(\pi /2)[\widehat{\mathrm{K}}_{-}+\widehat{\mathrm{K}}_{+}]}=e^{-i%
\widehat{\mathrm{K}}_{-}}e^{-i\widehat{\mathrm{K}}_{+}}e^{-i\widehat{\mathrm{%
K}}_{-}}=e^{-i\widehat{\mathrm{K}}_{+}}e^{-i\widehat{\mathrm{K}}_{-}}e^{-i%
\widehat{\mathrm{K}}_{+}}.  \label{fact3}
\end{gather}
They reproduce the two possible implementations of the Fourier transform by
optical elements, i.e.\ the $2f$-\textit{system} and the \textit{Fourier
tube}, respectively consisting of a single lens (of focal length~$f$) placed
midway between two reference planes separated by~$2f$ and of two identical
lenses (of focal length~$f$) separated by $f$. Both setups are described by
the ray-matrix $\mathbf{F}(f) = \binom{\,\,\,\,0\,\,\,\,\,\,\,\,\,f}{-1/f\,\,\,\,\,0}$; in particular, $f=1$ in~(\ref{fourier}) and so in~(\ref{fact3}). Unless otherwise specif\/ied, we will address to as Fourier
transformer the optical system described by the ray-matrix $\mathbf{F}(1)=%
\mathbf{F}=\binom{\,\,\,0\,\,\,\,\,1}{-1\,\,\,\,0}$.

Since every $\widehat{\mathrm{K}}\in \mathcal{G}$ is a linear combination of
the basis operators (\ref{su11}) and (\ref{weyl}), by (\ref{fact3}) also $\widehat{\mathcal{F}}\widehat{\mathrm{K}}\widehat{\mathcal{F}}^{-1}$and $\widehat{\mathcal{F}}^{-1}\widehat{\mathrm{K}}\widehat{\mathcal{F}}$ are
linear combinations of the same operators, and hence belong to~$\mathcal{G}$.

Then, as parallel to the eigenvalue problem for the operator~$\widehat{%
\mathrm{K}}$, signif\/ied by~(\ref{bas}) and~(\ref{alt}), we consider the
eigenvalue problem for the dual operator~$\widehat{\widetilde{\mathrm{K}}}$,
expressed by the equation
\begin{gather*}
\widehat{\widetilde{\mathrm{K}}}g_\lambda (\xi )=\lambda g_\lambda (\xi ),
%\label{ktfl}
\end{gather*}
for the f\/ixed-location operator, and by
\begin{gather*}
\widehat{\widetilde{\mathrm{K}}}(\zeta )w_\lambda (\xi ,\zeta )=\lambda
w_\lambda (\xi ,\zeta ),\qquad \widehat{\widetilde{\mathrm{K}}}(\zeta
)=e^{-i\zeta \widehat{\mathrm{K}}_{-}}\widehat{\widetilde{\mathrm{K}}}%
e^{i\zeta \widehat{\mathrm{K}}_{-}},  %\label{ktel}
\end{gather*}
for the relevant evolving-location operator.

Let us suppose that $\widehat{\widetilde{\mathrm{K}}}=\widehat{\mathcal{F}}%
\widehat{\mathrm{K}}\widehat{\mathcal{F}}^{-1}$. Accordingly, we see that
\[
g_\lambda =\widehat{\mathcal{F}}f_\lambda ,
\]
and hence the $w_\lambda $s are solutions of (\ref{normpar}) obtained by
propagating the Fourier transformed eigenfunctions of $\widehat{\mathrm{K}}$. Precisely, the $w_\lambda $s are obtained from the $v_\lambda $s through the
``local'' transformation
\begin{gather*}
w_\lambda (\xi ,\zeta )=e^{-i\zeta \widehat{\mathrm{K}}_{-}}\widehat{\mathcal{F}}e^{i\zeta \widehat{\mathrm{K}}_{-}}v_\lambda (\xi ,\zeta )=\widehat{\mathcal{A}}v_\lambda (\xi ,\zeta ),  %\label{wv}
\end{gather*}
by the symmetry operator
\begin{gather}
\widehat{\mathcal{A}}:=e^{-i\zeta \widehat{\mathrm{K}}_{-}}\widehat{\mathcal{%
F}}e^{i\zeta \widehat{\mathrm{K}}_{-}},  \label{appell}
\end{gather}
addressed to in \cite{torreat} as \textit{Appell transform operator}. In the
light of (\ref{hop}), it can be understood as a (back) evolving-location
Fourier operator: $\widehat{\mathcal{A}}=\widehat{\mathcal{F}}(\zeta )$. It
depends on the evolution variable $\zeta $; however, such a dependence will
not be explicitly displayed.

As noted in \cite{torreat}, the Appell operator (\ref{appell})
individualizes the optical ABCD system, which arises from the composition of
a Fourier transformer embedded between two free-space sections of length $%
-\zeta $ and $\zeta $. It is therefore described by the symplectic real
ray-matrix
\begin{gather}
\mathbf{M}_{\widehat{\mathcal{A}}}=
\begin{pmatrix}
1 & \zeta \\
0 & 1
\end{pmatrix}
\begin{pmatrix}
0 & 1 \\
-1 & 0
\end{pmatrix}
\begin{pmatrix}
1 & -\zeta \\
0 & 1
\end{pmatrix} =
\begin{pmatrix}
-\zeta & 1+\zeta ^2 \\
-1 & \zeta
\end{pmatrix} ,  \label{mappell}
\end{gather}
according to the $2\times 2$ matrix representation of each relevant optical
component (i.e.\ operator)\footnote{For the practical realization of such a system by
basic optical elements, as noted above, one may resort to free-sections and
lenses for both the free-propagation by~$\zeta $ and the Fourier
transformer. According to the analysis in~\cite{sudarshan}, the propagation
by the ``negative'' distance $-\zeta $ may be realized by free-sections and
lenses as well, specif\/ically by a sequence of three suitably designed lenses
separated by free-sections of proper lengths.}.

As a more explicit form of the Collins integral (\ref{uabcd}), the propagation of a signal through an ABCD system manifests in the transformation of the relevant wavefunction as  \cite{torrel,bandrespg}:
\begin{gather}
u(\xi ,\zeta _o)=\widehat{\mathcal{U}}_{\binom{A\,\,\,\,B}{C\,\,\,\,D}}u(\xi
,\zeta _i)=\frac 1{\sqrt{A}}e^{i\frac C{2A}\xi ^2}u\left(\frac \xi A,\zeta
_i+\frac BA\right).  \label{vwn}
\end{gather}
It signif\/ies that the whole action of the system is decomposed {\it{\`{a} la}} Wei--Norman \cite{wei} into a~sequence of a free-propagation by $B/A$, a lensing by focal power~$-C$, and a scaling by~$1/A$. In fact, $u(\xi ,\zeta _i+B/A)=[\widehat{\mathcal{U}}_{_{\sf{PWE}}}(\zeta _i+B/A)u_0](\xi )$ represents the  propagated form of the (ef\/fective or f\/ictitious) source function $u_0(\xi )$ with which the wavefunction $u(\xi ,\zeta _i)$ can be associated. Note that, since we are working with unitless variables, the of\/f-diagonal entries $B$ and $C$ (usually having the dimensions of length and 1/length) are intended to be normalized to~$b$ and~$1/b$.

Relation (\ref{vwn}) holds for $A\neq 0$. If $A=0$, according to (\ref{uabcd}%
) the wavefunction \mbox{transforms as}
\begin{gather}
\widehat{\mathcal{U}}_{\binom{0\,\,\,\,B}{C\,\,\,\,D}}u(\xi ,\zeta _i)=\frac
1{\sqrt{iB}}e^{i\frac D{2B}\xi ^2}\widetilde{u}\left(\frac \xi B,\zeta _i\right),
\label{vwn0}
\end{gather}
describing the ef\/fect of an optical Fourier transformer with focal length $f=B$ combined with a~modulation by the phase factor $\exp \left( i\frac
D{2B}\xi ^2\right) $.

By (\ref{vwn}) we can express the associated functions $w_\lambda $ in terms
of the $v_\lambda $s according to Appell's prescription:
\begin{gather}
w_\lambda (\xi ,\zeta )=\frac 1{\sqrt{i\zeta }}e^{i\frac{\xi ^2}{2\zeta }%
}v_\lambda \left(-\frac \xi \zeta ,-\frac 1\zeta \right).  \label{appell1}
\end{gather}

In the case of $\mathbf{M}_{\widehat{\mathcal{A}}}$, $A=0$ means $\zeta =0$,
which amounts to $D=0$ and $B=1$. Hence, following~(\ref{vwn0}) one recovers
the primary relation between the source functions: $w_\lambda (\xi ,0)= \widetilde{v}_\lambda (\xi ,0)$.

As noted in \cite{torreat}, relation (\ref{appell1}) can also be deduced by
acting on the initial function $v_0(\xi )$ by the operator $e^{-i\zeta
\widehat{\mathrm{K}}_{-}}\widehat{\mathcal{F}}$, and hence (apart from $%
\sqrt{i}$) through~(\ref{vwn}) with $\zeta _i=0$ and \mbox{$\binom{A\,\,\,\,B}{%
C\,\,\,\,D}\!=\!\binom{-\zeta \,\,\,1}{-1\,\,\,\,0}$.}

If in turn one has $\widehat{\widetilde{\mathrm{K}}}=\widehat{\mathcal{F}}%
^{-1}\widehat{\mathrm{K}}\widehat{\mathcal{F}}$, then
\begin{gather*}
w_\lambda (\xi ,\zeta )=e^{-i\zeta \widehat{\mathrm{K}}_{-}}\widehat{%
\mathcal{F}}^{-1}e^{i\zeta \widehat{\mathrm{K}}_{-}}v_\lambda (\xi ,\zeta ),
%\label{wvinv}
\end{gather*}
involving just the inverse of the Appell operator (\ref{appell}), i.e.
\begin{gather*}
\widehat{\mathcal{A}}^{-1}=e^{-i\zeta \widehat{\mathrm{K}}_{-}}\widehat{%
\mathcal{F}}^{-1}e^{i\zeta \widehat{\mathrm{K}}_{-}}=\widehat{\mathcal{F}}%
^{-1}(\zeta ).  %\label{appellinv}
\end{gather*}
It individualizes the ABCD system
\begin{gather}
\mathbf{M}_{\widehat{\mathcal{A}}^{-1}}=
\begin{pmatrix}
1 & \zeta \\
0 & 1
\end{pmatrix}
\begin{pmatrix}
0 & -1 \\
1 & 0
\end{pmatrix}
\begin{pmatrix}
1 & -\zeta \\
0 & 1
\end{pmatrix} =
\begin{pmatrix}
\zeta & -1-\zeta ^2 \\
1 & -\zeta
\end{pmatrix}
 =\mathbf{M}_{\widehat{\mathcal{A}}}^{-1},  \label{mappellinv}
\end{gather}
thus yielding the explicit transformation of the wavefunctions $v_\lambda $s
into the associated~$w_\lambda $s as
\begin{gather}
w_\lambda (\xi ,\zeta )=\frac 1{\sqrt{i\zeta }}e^{i\frac{\xi ^2}{2\zeta }%
}v_\lambda \left(\frac \xi \zeta ,-\frac 1\zeta \right).  \label{appell1inv}
\end{gather}

As a conclusion, we may say that in general the \textit{optical} Appell
transformation
\begin{gather}
w(\xi ,\zeta )=\frac 1{\sqrt{i\zeta }}e^{i\frac{\xi ^2}{2\zeta }}v\left(\pm \frac
\xi \zeta ,-\frac 1\zeta \right),  \label{appellgen}
\end{gather}
maps solutions of the linear 2D PWE into solutions \cite{torreat}. It is a
symmetry transformation for that equation, which traces back to a (direct or
inverse) Fourier relation between the source functions of the solutions it
connects. It manifests the action of the evolving-location Fourier operator
$\widehat{\mathcal{F}}(\zeta )$ or its inverse.

Note that the transformation (\ref{appellgen}) amounts to a mathematical
Fourier relation between the source functions of the involved solutions.
Indeed, an optical Fourier relation between the source functions would
demand for the trivial change of the phase factor as $e^{-i\pi
/4}\rightarrow e^{-i\pi /2}$.

As a basic example, we may consider the operators $\widehat{\mathrm{P}}$ and
$\widehat{\mathrm{X}}= \widehat{\mathcal{F}}\widehat{\mathrm{P}}\widehat{%
\mathcal{F}}^{-1}$. The eigenfunctions of $\widehat{\mathrm{P}}$, i.e.\ the
plane waves $f_\lambda (\xi )=\exp (i\lambda \xi )/\sqrt{2\pi }$, $
\lambda \in \mathbb{R}$, yield the wavefunctions
\begin{gather}
v_\lambda (\xi ,\zeta )=\tfrac 1{\sqrt{2\pi }}e^{i\lambda \xi }e^{-i\lambda
^2\zeta /2},  \label{chirp}
\end{gather}
which display the familiar frequency-chirping factor $\exp (-i\lambda
^2\zeta /2)$.

Applying the Appell transformation (\ref{appell1}) to (\ref{chirp}), one
just obtains the wavefunctions
\begin{gather}
w_\lambda (\xi ,\zeta )=\tfrac 1{\sqrt{2\pi i\zeta }}e^{i\frac{(\xi -\lambda
)^2}{2\zeta }},  \label{xwf}
\end{gather}
one would obtain by propagating the eigenfunctions $g_\lambda (\xi )=\delta
(\xi -\lambda )$, $\lambda \in \mathbb{R}$, of the operator $\widehat{\mathrm{X}}$.

Vice versa, the wavefunctions (\ref{xwf}) turn into the propagating plane
waves~(\ref{chirp}) by~(\ref{appell1inv}).

Another example of Appell pair of wavefunctions is provided by the two
kinds of Airy beams
\begin{gather*}
\psi _{\text{\textrm{KM}}}(\xi ,\zeta ,\lambda )=\tfrac 1{%
\sqrt{i\zeta }}e^{i\big(\frac 1{12\zeta ^3}+\frac{\xi ^2}{2\zeta }-\frac \xi
{2\zeta ^2}+\frac \lambda {2\zeta }\big)}\text{\textrm{Ai}}\left(\frac \xi \zeta
-\frac 1{4\zeta ^2}-\lambda \right),\nonumber\\
\psi _{\text{\textrm{BB}}}(\xi ,\zeta ,\lambda )=e^{-i\big(\frac{\zeta ^3}{12}-\frac{\zeta \xi }2+\frac \lambda 2\zeta \big)}\text{\textrm{Ai}}
\left(\xi -\frac{\zeta ^2}4-\lambda \right),\nonumber
\end{gather*}
\textrm{Ai} denoting the Airy function of the f\/irst kind~\cite{magnus}.
As seen in~\cite{torrea}, they are respectively obtained by propagating the
eigenfunctions of the operators $\widehat{\mathcal{P}}=2\widehat{\mathrm{K}}
_{+}+\widehat{\mathrm{P}}$ and $\widehat{\mathcal{X}}=2\widehat{\mathrm{K}}
_{-}+\widehat{\mathrm{X}}= \widehat{\mathcal{F}}\widehat{\mathcal{P}}%
\widehat{\mathcal{F}}^{-1}$. Accordingly, the former originate from the
cubic phase $\psi _{\text{\textrm{KM}}}(\xi ,0,\lambda )=\exp (i\lambda \xi
-i\xi ^3/3)/\sqrt{2\pi }$~\cite{kalnins1}, as the latter from an Airy
pattern: $\psi _{\text{\textrm{BB}}}(\xi ,0,\lambda )=\mathrm{Ai}(\xi
-\lambda )$~\cite{berry}; in both cases, $\lambda $ signif\/ies the eigenvalue
of the involved operator $\widehat{\mathcal{P}}$ or~$\widehat{\mathcal{X}}$.

It is easily verif\/ied that by (\ref{appell1}) $\psi _{\text{\textrm{KM}}%
}\rightarrow \psi _{\text{\textrm{BB}}}$, and vice versa by (\ref{appell1inv}%
) $\psi _{\text{\textrm{BB}}}\rightarrow \psi _{\text{\textrm{KM}}}$.

In the literature, the $\psi _{\text{\textrm{BB}}}$s are addressed to as
Airy beams. Along with other similar Airy-related solutions of the PWE, they
have attracted a great deal of interest (both analytical/numerical and
experimental), due to their non-spreading nature and their tendency to
freely accelerate; some pertinent titles are in \cite{berry,besieris,bandresa,siviloglou,besieris1,broky,morris,dai,baumgartl,ellenbogen,salandrino}.

\subsubsection{Fractional Appell transformation}\label{section2.1.3}

Conforming to the relevant generator $\widehat{\mathrm{K}}_{-}+\widehat{%
\mathrm{K}}_{+}$, the Fourier transform $\widehat{\mathcal{F}}$ can be
understood as the evolution operator describing the dynamics of a quantum
attractive oscillator (optically conveyed by a quadratic-index focusing
medium) observed at ``time'' $\tau =\pi /2$, the motion in the optical
phase-plane\footnote{Let us recall that the phase-plane (for systems with one degree
of freedom) is understood as the Cartesian plane formed by the relevant
canonically conjugate variables $(q,p)$, obeying the Poisson-bracket
relation $\{q,p\}=1$.

The light-ray coordinates $(q,p)$ of geometrical optics are canonically
conjugate variables. They span the optical phase-plane, where light rays are
represented by points, and accordingly the ray propagation through optical
systems corresponds to the ``motion'' of the relevant representative point.
For instance, free-sections and thin lenses produce under paraxial
propagation $q$ and $p$-shears in the phase-plane, i.e.\ translations of the
ray representative point respectively in the $q$ and $p$ direction.

Note that, in accord with the normalization of the space variables here
adopted, the ray-variables $(q,p)$ become $\binom \xi \upsilon \equiv \binom{%
q/w_0}{kw_0p}$, the ray optical momentum $p$, i.e.\ the angle relative to the
$z$-axis, being scaled to the natural far-f\/ield divergence $\vartheta
_\infty =1/kw_0$, associated with a beam having $w_0$ as a characteristic
width.} resulting in a clockwise rotation by $\pi /2$. Thereby, according to~(\ref{appell}), in the optical phase-plane the Appell transform amounts to a~$\pi /2$-clockwise rotation preceded and followed respectively by a negative
and positive $\xi $-shear.

As is well known, the Fourier transform $\widehat{\mathcal{F}}$ is a
specif\/ic determination of the more general fractional Fourier transform $\widehat{\mathcal{F}}^\alpha $, signif\/ied by the integral \cite{frt,ozaktas,lohmann,torrepo}
\begin{gather}
\big[\widehat{\mathcal{F}} ^\alpha \varphi \big](x):=\sqrt{\frac{1-i\cot
\phi }{2\pi }} \int_{-\infty }^{+\infty }dx'\, e^{\frac i{2\sin
\phi }(x'^2\cos \phi  + x^2\cos \phi  -2xx^{\prime })}\varphi
(x^{\prime }),  \label{falphaoperator2}
\end{gather}
the order $\alpha $ specifying the angle $\phi =\alpha \pi /2$, or
equivalently by the operator
\begin{gather*}
\widehat{\mathcal{F}}^\alpha =e^{i\alpha \pi /4}e^{-i\alpha (\pi /2)[
\widehat{\mathrm{K}}_{-}+\widehat{\mathrm{K}}_{+}]}=\big(\widehat{\mathcal{F}}
\big)^\alpha .  %\label{fft}
\end{gather*}
Again, the factor $e^{i\alpha \pi /4}$ relates the mathematical and optical
transforms, the latter being yielded by the Collins integral (\ref{uabcd})
in correspondence to the pertinent ray-matrix $\mathbf{F}^\alpha =\binom{\cos \phi \,\,\,\,\sin \phi }{-\sin \phi \,\,\,\,\cos \phi }$.

Both expressions reveal the periodicity
\begin{gather}
\widehat{\mathcal{F}}^{\alpha  + 4 j}=\widehat{\mathcal{F}}^\alpha
,\qquad j=0,\pm 1,\pm 2,\dots,  \nonumber
\end{gather}
so that the range of $\alpha $ can be limited to the interval $(-2,2]$.
Evidently, the ordinary transform~(\ref{fourier}) is recovered with $\alpha
=1$, i.e.\ $\phi =\pi /2$: $\widehat{\mathcal{F}}^1=\widehat{\mathcal{F}}$.
Also, $\widehat{\mathcal{F}}^0=\widehat{\mathrm{I}}$.

Paralleling $\widehat{\mathcal{F}}$, $\widehat{\mathcal{F}}^\alpha $ is
interpreted as the evolution operator associated with the harmonic
oscillator dynamics monitored continuously at ``time'' $\tau =\alpha \pi
/2=\phi $. Accordingly, the corresponding motion in the optical phase-plane
amounts to a clockwise rotation by~$\phi $.

As conveyed by (\ref{dis}), $\widehat{\mathcal{F}}^\alpha $ admits two
operator factorizations, which ref\/lect its optical realiza\-tions as
``fractional'' versions of the implementations of the ordinary transform.
One has, in fact, $\widehat{\mathcal{F}}^\alpha =e^{i\phi /2}e^{-i\tan (\phi
/2) \widehat{\mathrm{K}}_{-}}e^{-i\sin (\phi ) \widehat{\mathrm{K}}%
_{+}}e^{-i\tan (\phi /2) \widehat{\mathrm{K}}_{-}}$ for the single lens
realization (Lohmann type~I setup \cite{lohmann}) and  $\widehat{\mathcal{F}}^\alpha =e^{i\phi /2}e^{-i\tan
(\phi /2) \widehat{\mathrm{K}}_{+}}e^{-i\sin (\phi ) \widehat{\mathrm{K}}%
_{-}}e^{-i\tan (\phi /2) \widehat{\mathrm{K}}_{+}}$ for the two-lens
realization (Lohmann type II setup \cite{lohmann}). The former comprises a
thin lens of focal length $1/\sin (\phi )$ placed midway between two
reference planes spaced by $2\tan (\phi /2)$, whereas the latter realizes
the dual conf\/iguration, composed by two thin lenses of focal length $\cot
(\phi /2)$ separated by~$\sin (\phi )$.

It has been suggested in \cite{torreat} to consider the \textit{fractional}
\textit{Appell transform operator} $\widehat{\mathcal{A}}^\alpha $, which
then, as fractional version of the operator $\widehat{\mathcal{A}}$, has
been written in the form
\begin{gather}
\widehat{\mathcal{A}}^\alpha :=e^{-i\zeta \widehat{\mathrm{K}}_{-}}\widehat{%
\mathcal{F}}^\alpha e^{i\zeta \widehat{\mathrm{K}}_{-}}=\widehat{\mathcal{F}}%
^\alpha (\zeta ),  \label{appellf}
\end{gather}
whose dependence on $\zeta $ will not explicitly displayed. The ordinary
operator (\ref{appell}) corresponds to $\alpha =1$, whilst the inverse is $(%
\widehat{\mathcal{A}}^\alpha )^{-1}=\widehat{\mathcal{A}}^{-\alpha }$, since
$(\widehat{\mathcal{F}}^\alpha )^{-1}=\widehat{\mathcal{F}}^{-\alpha }$.

The operator (\ref{appellf}) individualizes the ABCD system, described by
the matrix
\begin{gather*}
\mathbf{M}_{\widehat{\mathcal{A}}^\alpha }=
\begin{pmatrix}
1 & \zeta \\
0 & 1
\end{pmatrix}
\begin{pmatrix}
\cos \phi & \sin \phi \\
-\sin \phi & \cos \phi
\end{pmatrix}
\begin{pmatrix}
1 & -\zeta \\
0 & 1
\end{pmatrix}
  =
\begin{pmatrix}
\cos \phi -\zeta \sin \phi & (1+\zeta ^2)\sin \phi \\
-\sin \phi & \cos \phi +\zeta \sin \phi
\end{pmatrix}
 ,  %\label{mappellf}
\end{gather*}
which reproduces the matrices (\ref{mappell}) for $\phi =\pi /2$ and (\ref{mappellinv}) for $\phi =-\pi /2$.

{\sloppy Then, according to (\ref{vwn}) and (\ref{vwn0}), we see that the \textit{fractional Appell transformation} amounts to the replacement of the
wavefunction $v(\xi ,\zeta )$ by the wavefunction $w(\xi ,\zeta )$ given by~\cite{torreat}
\begin{gather}
w(\xi ,\zeta )=\frac{e^{i\phi /2}}{\sqrt{\cos \phi -\zeta \sin \phi }}e^{-i%
\frac{\xi ^2\sin \phi }{2(\cos \phi -\zeta \sin \phi )}}v\left(\frac \xi {\cos
\phi -\zeta \sin \phi },\frac{\sin \phi +\zeta \cos \phi }{\cos \phi -\zeta
\sin \phi }\right),  \label{wffat}
\end{gather}
as far as $\cos \phi -\zeta \sin \phi \neq 0$, whilst in the case when $\cos
\phi -\zeta \sin \phi =0$ (i.e. $\zeta =\cot \phi $, which can occur for $%
\phi \in (-\pi ,-\pi /2]\cup [0,\pi /2]$), it acts as
\begin{gather*}
w(\xi ,\zeta )=\frac{e^{i(\phi -\pi /2)/2}}{\sqrt{(1+\zeta ^2)\sin \phi }}%
e^{i\frac \zeta {1+\zeta ^2}\xi ^2}\widetilde{v}\left(\frac \xi {(1+\zeta ^2)\sin
\phi },\zeta \right) \qquad \text{for}\quad \zeta =\cot \phi .
%\label{wffat0}
\end{gather*}

}

The fractional Appell transformation identif\/ies a family of transformations
parameterized by the continuous parameter $\phi \in (-\pi ,\pi ]$. It would
allow one to pass, for instance, from the $\psi _{\text{\textrm{KM}}}$s to
the $\psi _{\text{\textrm{BB}}}$s through a continuum of ``states'' f\/illing
up the direct transformation $\psi _{\text{\textrm{KM}}}\rightarrow \psi _{\text{\textrm{BB}}}$, considered before, occurring in fact for $\phi =\pi /2$~\cite{torreat}.

We conclude by recalling that, as a consequence of the properties of the
fractional Fourier transform, the fractional Appell operator satisf\/ies the
basic properties:
\begin{enumerate}\itemsep=0pt
\item[$(i)$]  $\widehat{\mathcal{A}}^\alpha $
is continuous for all values of the order parameter $\alpha $,

\item[$(ii)$] $\widehat{\mathcal{A}}^\alpha $ obeys the group property, so
that composing two operators of order $\alpha _{_1}$ and $\alpha _{_2}$
yields the operator of order $\alpha _{_1}$ $+$ $\alpha _{_2}$:
\[
\widehat{\mathcal{A}}^{\alpha _1}\widehat{\mathcal{A}}^{\alpha _2}=\widehat{%
\mathcal{A}}^{\alpha _1+\alpha _2}=\widehat{\mathcal{A}}^{\alpha _2}\widehat{%
\mathcal{A}}^{\alpha _1},
\]

\item[$(iii)$] $\widehat{\mathcal{A}} ^\alpha $ reduces to the ordinary
operator for $\alpha =1$: $\widehat{\mathcal{A}}^1=\widehat{\mathcal{A}}$,
and the identity operator for $\alpha =0$: $\widehat{\mathcal{A}}^0=\widehat{\mathrm{I}}$.
\end{enumerate}

The property $(i)$ follows from the continuity of the Collins integral
(\ref{uabcd}) on account of that of the ray-matrix entries as functions of $\phi $ (or, $\alpha $). Property $(ii)$ ensues from the fact that the
composition of the propagation integrals~(\ref{uabcd}) ref\/lects that of the
ray-matrices, and hence $\widehat{\mathcal{A}}^{\alpha _1}\widehat{\mathcal{A%
}}^{\alpha _2}\leftrightarrow \mathbf{M}_{\widehat{\mathcal{A}}^{\alpha _1}}%
\mathbf{M}_{\widehat{\mathcal{A}}^{\alpha _2}}=\mathbf{M}_{\widehat{\mathcal{%
A}}^{\alpha _1+\alpha _2}}\leftrightarrow \widehat{\mathcal{A}}^{\alpha
_1+\alpha _2}$ in accord with the addition formulae of the circular
functions. Property $(iii)$ is implied by that $\widehat{\mathcal{F}}^1=\widehat{\mathcal{F}}$ and $\widehat{\mathcal{F}}^0=\widehat{\mathrm{I}}$. In addition, as for $\widehat{\mathcal{F}}^\alpha $, the group property
$(ii)$ allows $\widehat{\mathcal{A}}^\alpha $ to be mathematically
understood as the $\alpha $-th power of $\widehat{\mathcal{A}}$.

Evidently, the expressions (\ref{appellgen}) and (\ref{wffat}) conform to
the well-known rule, which, according to standard results from Lie theory,
conveys the action on functions in $\mathfrak{F}$ of any operator in the PWE
symmetry group spanned by $\big\{ \widehat{\mathrm{K}}_{+},\widehat{\mathrm{K}}_3,\widehat{\mathrm{K}}_{-}\big\} $~\cite{kalnins1,miller}. As earlier
noted, the analysis above favors the ``visualization'' of the Appell
transformation in optical terms, and makes also explicit the connection
between some given transformations, which so appear as its ``fractional
powers''.

\subsection{2D ``radial''  paraxial wave equation}\label{section2.2}

As we know, the 3D PWE in (normalized) circular cylindrical coordinates $%
(\rho ,\varphi ,\zeta )$,
\begin{gather*}
\left[ 2i\frac \partial {\partial \zeta }+\frac{\partial ^2}{\partial \rho
^2}+\frac 1\rho \frac \partial {\partial \rho }+\frac 1{\rho ^2}\frac{%
\partial ^2}{\partial \varphi ^2}\right]u(\rho ,\varphi ,\zeta )=0,
\end{gather*}
with $\rho =\sqrt{\xi ^2+\eta ^2}$ and $\varphi =\arctan (\eta /\xi )$,
allows for separable-variable solutions as
\begin{gather}
u(\rho ,\varphi ,\zeta )=\Phi (\rho ,\zeta )e^{im\varphi}.  \label{factcc}
\end{gather}
The evolution of the radial wavefunction $\Phi (\rho ,\zeta )$ for a given
azimuthal index $m$ is accordingly ruled by the 2D PWE in the radial
coordinate $\rho $:
\begin{gather}
\left[ 2i\frac \partial {\partial \zeta }+\frac{\partial ^2}{\partial \rho
^2}+\frac 1\rho \frac \partial {\partial \rho }-\frac{m^2}{\rho ^2}\right]\Phi
(\rho ,\zeta )=0.  \label{parrad}
\end{gather}

In order to apply the above illustrated procedure to this equation, we
resort to an appropriate representation of the $sl(2,\mathbb{R})\simeq su(1,1)$ generators
for the symmetry of concern, namely
\begin{gather}
\widehat{\mathcal{K}}_{+}:=\frac 12\rho ^2,\qquad \widehat{\mathcal{K}}%
_{-}:=-\frac 12\left(\frac{\partial ^2}{\partial \rho ^2}+\frac 1\rho \frac
\partial {\partial \rho }-\frac{m^2}{\rho ^2}\right),\qquad \widehat{\mathcal{K}}%
_3:=-\frac i2\left(\rho \frac \partial {\partial \rho }+1\right).  \label{su1c}
\end{gather}
As their planar counterpart (\ref{su11}), the above relate to the operators $
\widehat{\mathbf{x}}^2$, $\widehat{\mathbf{p}}^2$ and $\widehat{\mathbf{x}}
\cdot \widehat{\mathbf{p}}$, properly expressed for a circular cylindrical
symmetry.

The relevant propagator $\widehat{\mathcal{U}}_{_{\mathsf{PWE}}}(\zeta
):=\exp (-i\zeta  \widehat{\mathcal{K}}_{-})$ amounts to a Hankel-like
transform as that in rectangular coordinate amounts to the Fresnel transform~(\ref{hfi}). One has in fact:
\begin{gather}
\Phi (\rho ,\zeta )=\big[\widehat{\mathcal{U}}_{_{\mathsf{PWE}}}(\zeta )f\big](\rho )=%
\tfrac{(-i)^{m+1}}\zeta \int_0^\infty e^{\frac i{2\zeta }(\rho ^{\prime
}{}^2+\rho ^2)}J_m\big(\tfrac{\rho \rho ^{\prime }}\zeta \big)f(\rho ^{\prime })\rho
^{\prime }d\rho ^{\prime },  \label{prad}
\end{gather}
$J_m$ denoting the Bessel function of the f\/irst kind and order $m$ \cite
{magnus}. Of course, $\Phi (\rho ,0)=f(\rho )$.

The above is a particular form of the (real or complex) radial canonical
transform \cite{siegman}
\begin{gather}
\Phi (\rho ,\zeta _o)=\widehat{\mathcal{U}}_{\binom{A\,\,\,\,B}{C\,\,\,\,D}%
}\Phi (\rho ,\zeta _i)=\tfrac{(-i)^{m+1}}B\int_0^{+\infty }e^{\frac
i{2B}(A\rho ^{\prime }{}^2+D\rho ^2)}J_m\big(\tfrac{\rho \rho ^{\prime }}B\big)\Phi
(\rho ^{\prime },\zeta _i)\rho ^{\prime }d\rho ^{\prime },  \label{uabcdr}
\end{gather}
corresponding to the canonical transformation conveyed by the (real or
complex) symplectic matrix $\mathbf{M}=\binom{A\,\,\,\,B}{C\,\,\,\,D}$. It
basically follows from the 2D Collins integral under the circular
cylindrical symmetry assumption (\ref{factcc}). Hence, it relates the radial
wavefunctions of the wavef\/ields (of a~given azimuthal symmetry) at the input
and output planes at~$\zeta _i$ and $\zeta _o$, between which the optical
system, described by the ray-matrix~$\mathbf{M}$, is conventionally intended
to operate.

As (\ref{uabcd}), when $A\neq 0$ the dif\/fraction integral (\ref{uabcdr}) can be seen
as resulting from a free propagation by $B/A$ , followed by a lensing with
focal power $-C$ and a scaling by $1/A$, so that
\begin{gather}
\Phi (\rho ,\zeta _o)=\widehat{\mathcal{U}}_{\binom{A\,\,\,\,B}{C\,\,\,\,D}%
}\Phi (\rho ,\zeta _i)=\frac 1Ae^{i\frac C{2A}\rho ^2}\Phi \left(\frac \rho
A,\zeta _i+\frac BA\right).  \label{vwncc}
\end{gather}
Here  $\Phi (\upsilon ,\zeta _i+B/A)=[\widehat{\mathcal{U}}_{_{\sf{PWE}}}(\zeta _i+B/A)\Psi _0](\upsilon )$ represents  the radial function composing the propagated form of the (ef\/fective or
f\/ictitious) source function $u(\upsilon ,\varphi ,0)=$ $\Psi _0(\upsilon
)e^{im\varphi }$, with which the wavefunction $u(\upsilon ,\varphi ,\zeta
_i)=\Psi (\upsilon ,\zeta _i)e^{im\varphi }$ can be associated and whose azimuthal symmetry, accounted for by the factor $e^{im\varphi }$ and supposedly
preserved by the propagation, is implicitly conveyed by the dependence of the free-propagation integral (\ref{prad})  on
the azimuthal index $m$.

In analogy with (\ref{vwn0}), the case $A=0$ yields the transformation
\begin{gather}
\widehat{\mathcal{U}}_{\binom{0\,\,\,\,B}{C\,\,\,\,D}}\Psi (\rho ,\zeta _i)=%
\frac{(-i)^{m+1}}Be^{\frac{iD}{2B}\rho ^2}\widetilde{\Psi }\left(\frac \rho
B,\zeta _i\right),  \label{vwncc0}
\end{gather}
involving, as expected, the Hankel transform $\widetilde{\Psi }(\rho
/B,\zeta _i)$ of order $m$ of the radial wavefunction at the input plane,
modulated by the phase factor $\exp \left( i\frac D{2B}\rho ^2\right) $.

As is well known, in fact, the 2D Fourier transform of a function $v(\rho
,\varphi )$ obeying the factori\-za\-tion (\ref{factcc}) with respect to the
polar coordinates, i.e. $v(\rho ,\varphi )=e^{im\varphi }\Phi (\rho )$,
turns into
\[
\widehat{\mathcal{F}}_\xi \widehat{\mathcal{F}}_\eta u(\rho ,\varphi
)=e^{im(\varphi -\pi /2)}\widetilde{\Phi }(\rho ),
\]
with $\widetilde{\Phi }(\rho )$ signifying the Hankel transform of $\Phi
(\rho )$ of order $m$, which according to the usual def\/inition means
\begin{gather}
\widetilde{\Phi }(\rho )=\big[\widehat{\mathcal{H}}_m\Phi \big](\rho
):=\int_0^\infty J_m(\rho \rho ^{\prime })\Phi (\rho ^{\prime })\rho
^{\prime }d\rho ^{\prime }.  \label{hankel}
\end{gather}

As for (\ref{normpar}), several solutions of (\ref{parrad}) have been
identif\/ied, which can be understood as arising from eigenstates of def\/inite
operators in the algebra generated by $\{\widehat{\mathcal{K}}_{+},\widehat{%
\mathcal{K}}_3,\widehat{\mathcal{K}}_{-}\}$ \cite{kalnins1,torresp}. Thus,
for instance, the eigenfunctions of the operator $\widehat{\mathcal{K}}_{\xi
_0}= \widehat{\mathcal{K}}_3-(2/\xi _0)\widehat{\mathcal{K}}_{+}$ evolve
into wavefunctions, which, as the aforementioned Weber--Hermite solutions of~(\ref{normpar}), depend on three independent parameters and comprise a
complex quadratic exponential modulated by the Whittaker f\/irst function $M_{\kappa ,\mu }$ of suitable argument~\cite{torresp}. As remarked, such
solutions have been also deduced in~\cite{bandresr} through an appropriate
variable-separation ansatz.

\subsubsection{The Hankel transform and the optical (radial) Appell
transformation}\label{section2.2.1}

In full analogy with the Fourier transform, the Hankel transform (\ref{hankel}) of order $m$ can be given an operator representation in terms of
the algebra generators $\widehat{\mathcal{K}}_{+}$ and $\widehat{%
\mathcal{K}}_{-}$; namely,
\begin{gather}
\widehat{\mathcal{H}}_m=i^{m+1}e^{-i(\pi /2)[\widehat{\mathcal{K}}_{-}+%
\widehat{\mathcal{K}}_{+}]}.  \label{facthankel}
\end{gather}
As before, the factor $i^{m+1}$ allows the mathematical transform (\ref{hankel}) to be matched to the optical transform, conveyed by the
Huygens--Hankel integral (\ref{uabcdr}) for the Fourier matrix $\mathbf{F}$.

Then, going through the same procedure as before, we can see that the radial
Appell transformation (for a given azimuthal index $m$) is described by the
operators
\begin{gather*}
\widehat{\mathcal{A}}_m:=e^{-i\zeta \widehat{\mathcal{K}}_{-}}\widehat{%
\mathcal{H}}_me^{i\zeta \widehat{\mathcal{K}}_{-}}\qquad \text{and} \qquad \widehat{\mathcal{A}}_m^{-1}:=e^{-i\zeta \widehat{\mathcal{K}}%
_{-}}\widehat{\mathcal{H}}_m^{-1}e^{i\zeta \widehat{\mathcal{K}}_{-}},
%\label{appellr}
\end{gather*}
whose dependence on $\zeta $ will not explicitly displayed.

They respectively amount to the ABCD matrices (\ref{mappell}) and (\ref
{mappellinv}), and hence, by (\ref{vwncc}), to the (radial) wavefunction
transformations (for a given azimuthal index $m$):
\begin{gather}
\Psi (\rho ,\zeta )=\frac{(\pm i)^m}{i\zeta }e^{i\frac{\rho ^2}{2\zeta }%
}\Phi \left(\mp \frac \rho \zeta ,-\frac 1\zeta \right).  \label{appradi}
\end{gather}
In accord with (\ref{vwncc0}), accounting for the case $A=0$, we recover the
Hankel transform-relation between the source functions$:\Psi (\rho ,0)=\widetilde{\Phi }(\rho ,0)$.

The two possibilities conveyed by (\ref{appradi}) actually yield the same
expression for the transformed wavefunction; this ref\/lects the fact that $%
\widehat{\mathcal{A}}_m=$\thinspace \thinspace $\widehat{\mathcal{A}}_m^{-1}$
as a consequence of the self-reciprocity of the Hankel transform: $\widehat{%
\mathcal{H}}_m=\widehat{\mathcal{H}}_m^{-1}$.

In analogy with the ``linear'' operator $\widehat{\mathcal{A}}$, $\widehat{\mathcal{A}}_m$ can be regarded as a (back) evolving-location Hankel
transform operator: $\widehat{\mathcal{A}}_m=\widehat{\mathcal{H}}_m(\zeta )$.

Paralleling the cases discussed in Section~\ref{section2.1.2}, we consider, as an Appell pair
of solutions of~(\ref{parrad}), the Bessel beams \cite{durnin} and the
Bessel--Gauss beams \cite{sheppardbg}. The former can be interpreted as
propagated forms of the eigenfunctions $J_m(\lambda \rho )$, $\lambda
\in \mathbb{R}$, of the free-Hamiltonian operator $\widehat{\mathcal{K}}_{-}$:{\samepage
\begin{gather*}
B_{\lambda ,m}(\rho ,\zeta )=e^{-i\lambda ^2\zeta /2}J_m(\lambda \rho ),
\end{gather*}
the eigenvalue $\lambda $ signifying the transverse component of the
wavenumber: $\lambda \leftrightarrow k_{\perp }$ \cite{kalnins1,torresp}.}

By the Appell transformation (\ref{appradi}), the $B_{\lambda ,m}$s turn
into the Bessel--Gauss beams:
\begin{gather*}
BG_{\lambda ,m}(\rho ,\zeta )=\frac{(-i)^{m+1}}\zeta e^{i\frac{\lambda ^2}{%
2\zeta }}e^{i\frac{\rho ^2}{2\zeta }}J_m\big(\lambda \tfrac \rho \zeta \big),
\end{gather*}
which in turn arise from the eigenfunctions $\delta (\rho -\lambda )$, $\lambda \in \mathbb{R}$, of the dual operator $\widehat{\mathcal{K}}_{+}$
\cite{kalnins1,torresp}.

As is well known, the Bessel modes are dif\/fractionless (as the Airy beams $%
\psi _{\text{\textrm{BB}}}$\footnote{Note that the comparison should more correctly involve
the eigenfunctions of $\widehat{\mathrm{P}}$, i.e.\ the plane waves $%
e^{i\lambda \xi }$.}), whilst the Bessel--Gauss modes, which as seen evolve
from the Hankel transform of the source functions of the former, have
complementary properties in both the space and spatial frequency domains.

Evidently, applying the radial fractional Appell transformation (for the
given~$m$), plainly understood as
\begin{gather*}
\widehat{\mathcal{A}}_m^\alpha :=e^{-i\zeta \widehat{\mathcal{K}}_{-}}%
\widehat{\mathcal{H}}_m^\alpha e^{i\zeta \widehat{\mathcal{K}}_{-}}=\widehat{%
\mathcal{H}}_m^\alpha (\zeta ),  %\label{appellrf}
\end{gather*}
one could follow the continuous transformation from the Bessel to the
Bessel--Gauss modes at any $\zeta $. Here, $\widehat{\mathcal{H}}_m^\alpha $
signif\/ies the fractional Hankel transform, which, resorting to the
fractional Fourier transform matrix $\mathbf{F}^\alpha $, by (\ref{uabcdr})
means
\begin{gather}
\big[ \widehat{\mathcal{H}}_m^\alpha \Phi \big](\rho )=\big[e^{i(m+1)\phi
}e^{-i\phi [\widehat{\mathcal{K}}_{-}+\widehat{\mathcal{K}}_{+}]}\Phi \big](\rho
)\nonumber\\
\hphantom{\big[ \widehat{\mathcal{H}}_m^\alpha \Phi \big](\rho )}{} :=\tfrac{e^{i(m+1)(\phi -\pi /2)}}{\sin \phi }\int_0^\infty e^{\frac{i\cos
\phi }{2\sin \phi }(\rho ^{\prime }{}^2+\rho ^2)}J_m\big(\tfrac{\rho \rho
^{\prime }}{\sin \phi }\big)f(\rho ^{\prime })\rho ^{\prime }d\rho ^{\prime
},   \label{hfr}
\end{gather}
with $\phi =\alpha \pi /2\in (-\pi ,\pi ]$. Therefore, as far as $\cos \phi
-\zeta \sin \phi \neq 0$, $\widehat{\mathcal{A}}_m^\alpha $ yields the
transformation
\begin{gather*}
\Psi (\rho ,\zeta )=\frac{e^{i(m+1)\phi }}{\cos \phi -\zeta \sin \phi }e^{-i%
\frac{\rho ^2\sin \phi }{2(\cos \phi -\zeta \sin \phi )}}\Phi \left(\frac \rho
{\cos \phi -\zeta \sin \phi },\frac{\sin \phi +\zeta \cos \phi }{\cos \phi
-\zeta \sin \phi }\right),  %\label{atrfr}
\end{gather*}
whilst for $\zeta =\cot \phi $ it acts as
\begin{gather*}
\Psi (\rho ,\zeta )=\frac{e^{i(m+1)(\phi -\pi /2)}}{(1+\zeta ^2)\sin \phi }%
e^{i\frac \zeta {1+\zeta ^2}\rho ^2}\widetilde{\Phi }\left(\frac \rho {(1+\zeta
^2)\sin \phi },\zeta \right) \qquad \text{for}\quad \zeta =\cot \phi .
%\label{atrfr0}
\end{gather*}

\subsection{Optical Appell transformation and duality.\\
Self-Fourier/self-Hankel and self-Appell wavefunctions}\label{section2.3}

Since, as reviewed above, the Appell transformation manifests the
correspondence between wavefunctions generated by Fourier or Hankel pairs of
functions \cite{torreat}, it naturally relates to the concept of beam
duality \cite{lohmannd,sheppard}. However, it does not connect
wavefunctions, which are locally dual, but connects wavefunctions whose
duality traces back to the respective source functions.

An ``Appell transformer''~-- provided it be implementable~-- would be so an
optical device turning a wavefunction into that which one would generate by
propagating the Fourier transform of the source function of the original
wavefunction. In a sense, as the Fourier transformer can be understood as a
``local dual switch'', the ``Appell transformer'' might be understood as an
``initial-plane dual switch''. Correspondingly, the \textit{fractional
Appell transformations} individualize a family of symmetry transformations
parameterized by a continuous parameter, which, f\/illing the gap between the
``evolution-lines'' of a function and its Fourier/Hankel transform, might be
seen as relating to a sort of ``fractional beam duality''.

In the light of the examples considered above, i.e.\ Airy and Bessel beams,
we may visualize the problem in terms of two ideal paths, running parallel
to each other to depict the $\zeta $-lines along which the evolution of a
given source function and of its Fourier or Hankel transform occur. The
Appell transformation connects one path to the other at any desired $\zeta $
or better one path to any other path between the two ``extreme'' ones, thus
allowing us to ``have a look'' at or to turn the wavefunction at hand into a
wavefunction whose properties in the space and spatial frequency domains are
a desired mixture of those of the wavefunctions ``lying'' on the two extreme
paths.

Evidently, when the source functions are self-Fourier or self-Hankel
functions, the two paths collapse one into the other. The Appell
transformation in fact comes to reproduce at any $\zeta $ the direct (or,
inverse) self-Fourier/self-Hankel relations obeyed by the source functions.
This is the case, for instance, of the standard Hermite--Gauss and
Laguerre--Gauss modes, which respectively arise from the eigenfunctions of
the self-dual Lie operators $\widehat{\mathrm{K}}_{-}+\widehat{\mathrm{K}}%
_{+}$ and $\widehat{\mathcal{K}}_{-}+\widehat{\mathcal{K}}_{+}$; as seen
above, the latter are the generators of the (both ordinary and fractional)
Fourier and Hankel transforms.

In fact, taking into account the explicit expressions for the quoted
(normalized) modes, i.e.
\[
sHG_n(\xi ,\zeta )=\frac 1{\sqrt{2^nn!\mu (\zeta )\sqrt{\pi }}}\left( \frac{%
\mu ^{*}(\zeta )}{\mu (\zeta )}\right) ^{n/2}e^{-\frac{\xi ^2}{2\mu (\zeta )}%
}H_n\left(\frac \xi {|\mu (\zeta )|}\right),
\]
for the standard Hermite--Gauss modes, and
\[
sLG_{n,m}(\rho ,\zeta )=\sqrt{\frac{2n!}{(n+m)!}}\frac 1{\mu (\zeta
)^{m+1}}\left( \frac{\mu ^{*}(\zeta )}{\mu (\zeta )}\right) ^n\rho ^me^{-
\frac{\rho ^2}{2\mu (\zeta )}}L_n^m\left(\frac{\rho ^2}{|\mu (\zeta )|^2}\right),
\]
for the standard Laguerre--Gauss modes, where
\[
\mu (\zeta )=1+i\zeta ,
\]
$H_n$ denotes the Hermite polynomial of degree $n$ and $L_n^m$ the
generalized Laguerre polynomial of degree $n$ and order $m$ \cite{magnus},
we can verify that
\begin{gather*}
\widehat{\mathcal{A}}^\alpha sHG_n(\xi ,\zeta )=(-i)^{\alpha
n}sHG_n(\xi ,\zeta ), \\
\widehat{\mathcal{A}}_m^\alpha sLG_{n,m}(\rho ,\zeta
)=(-1)^{\alpha n}sLG_{n,m}(\rho ,\zeta ).
\end{gather*}
The above just reproduce the relations holding between the relevant source
functions, which are the Hermite--Gauss and Laguerre--Gauss functions $%
sHG_n(\xi ,0)=\frac 1{\sqrt{2^nn!\sqrt{\pi }}}e^{-\xi ^2/2}H_n(\xi )$ and $%
sLG_{n,m}(\rho ,0)=\sqrt{\frac{2n!}{(n+m)!}}\rho ^me^{-\rho ^2/2}L_n^m(\rho
^2)$, and the respective Fourier and Hankel transforms.

In a sense, the standard Hermite--Gauss and Laguerre--Gauss modes can be
considered as self-Appell wavefunctions, respectively suitable to a
rectangular and circular cylindrical geometry.

\section{Canonical transforms: a short review}\label{section3}

As seen, the interpretation of the Appell transformation within the optical
context naturally resorts to the Fourier and Hankel transforms \cite{torreat}, which ultimately are integral transforms.

In general, integral transforms provide a well established and valuable
method to solve problems in several areas of both physics and applied
mathematics. As we know, the roots of the method can be traced back to the
original work by Oliver Heaviside on the ordinary dif\/ferential equations
with constant coef\/f\/icients occurring in the theory of electric circuits \cite{heaviside1,heaviside2}. Due also to the availability of large scale
computers, the method has then been increasingly extended to a wide range of
physical and mathematical problems, basically described by partial
dif\/ferential equations with assigned boundary and initial conditions~\cite{tranter,davies,wolfb}.

In particular, a special class of integral transforms, named \textit{canonical transforms}~(CT), appear widely in optics, in electromagnetism, in
classical and quantum mechanics as well as in computational and applied
mathematics. Their link to the canonical transformations and to the
parabolic dif\/ferential equations as well as to the theory of special
functions has been deeply analyzed. Fourier, bilateral Laplace, Bargmann,
Weierstrass--Gauss transforms as well as Hankel and Barut--Girardello
transforms are examples of CTs.

Canonical transformations play a crucial role in classical mechanics. When
applying the inherent formalism to quantum mechanics, the CTs naturally
arise (under specif\/ic conditions) as associated representations (unitary or
not) acting between suitably constructed Hilbert spaces of functions on the
real or complex domain~\cite{bargmann,moshinsky,quesne,wolfl,wolfr,kramer}.

On the other hand, the initial value problem for evolutionary equations is
usually formulated in terms of the evolution (or, displacement) operator.
The 2D PWE and the (1+1)D SE are examples of evolution equations, whose
displacement operators realize a unitary mapping of the Hilbert space of
square integrable functions into itself. The same can be said for the 1D~HE
although in this case the relevant evolution operator no longer generates a
unitary mapping. It is proved that CTs can be realized as evolution
operators, generated by second-order dif\/ferential operators through
exponentiation to the group by real or complex parameters, and hence as such
can be associated with evolution equations, ruled by Hamiltonian-like
operators (not necessarily Hermitian) which are quadratic in the inherent
canonically conjugate variables \cite{wolfpde}.

In addition, CTs directly relate to some aspects of the theory of special
functions through the eigenvalue problem~\cite{wolfef}. In fact, special
functions, like the Hermite--Gauss, the Laguerre--Gauss and the parabolic
cylinder functions, are self-reciprocal under some of the aforementioned CTs.

The Fourier and Hankel transforms are basic examples of CTs. Their role has
been enlarged to more general contexts by the introduction of the
corresponding transforms of fractional order~\cite{namias1,namias2,mcbride},
which, for instance, in the f\/ield of optics~\cite{frt,ozaktas,lohmann,torrepo}, gave
rise to a great variety of applications, investigations and new formulations
in an increasingly enriched optics scenario, that in turn stimulated further
general analyses of the linear CTs from both theoretical and
applicative/numerical points of view. Few recent titles are in \cite{pei,torretr,alieva,stern,deng,koc}.

The Fourier and Hankel transforms are real (respectively, linear and radial)
CTs. In Section~\ref{section4.2} we will see that the \textit{caloric} Appell
transformation, relevant to the 1D~HE, naturally relates to the bilateral
Laplace transform, which in contrast is a special type of complex (linear)
CTs.

\subsection{Linear canonical transforms}\label{section3.1}

We recall that real linear\footnote{The term \textit{linear} accounts for the integration involving
the real line $\mathbb{R}$.} CTs realize a unitary mapping of the Hilbert space $\mathfrak{L}^2(\mathbb{R})$ into itself through the integral transform \cite{moshinsky,quesne,wolfl,wolfr}
\begin{gather}
\widehat{\mathsf{T}}_{{}_{\mathbf{M}}}:\varphi \in \mathfrak{L}^2(\mathbb{R}%
)\rightarrow [\widehat{\mathsf{T}}_{{}_{\mathbf{M}}}\varphi
](x):=\int_{-\infty }^\infty \mathsf{K}_{{}_{\mathbf{M}}}(x,x^{\prime
})\varphi (x^{\prime })dx^{\prime }\in \mathfrak{L}^2(\mathbb{R}),  \label{can2}
\end{gather}
whose kernel
\begin{gather}
\mathsf{K}_{{}_{\mathbf{M}}}(x,x^{\prime }):=\frac 1{\sqrt{2\pi iB}}e^{\frac
i{2B}(Ax^{\prime 2}+Dx^2-2xx^{\prime })}  \label{can3}
\end{gather}
depends on the three linearly independent entries $A$, $B$, $C$ of
a $2\times 2$ unimodular real matrix $\mathbf{M}=
\left( _{C\,\,D}^{A\,\,\,B}\right) $, with the aforestated convention on the square root of a (complex) number and the requirement $A/B\geq 0$ resulting form the integrability condition. The matrix $%
\mathbf{M}$ belongs to the three-parameter real symplectic group $Sp(2,\mathbb{R%
})\simeq SL(2,\mathbb{R})$, and in practice specif\/ies the canonical
transformation of the involved (classical or quantum) canonically conjugate variables\footnote{As is well known, classically a canonical
transformation is a change of the phase-space variables $(q,p)$ $\rightarrow
$ $(q^{\prime }(q,p),p^{\prime }(q,p))$, which preserves the Poisson bracket
$\{q,p\}=1=\left\{ q^{\prime },p^{\prime }\right\} $. As a straightforward
extension of the above concept, a quantum canonical transformation is
def\/ined as a change of the conjugate non-commuting observables $\widehat{q}%
\rightarrow $ $\widehat{q}^{\prime }(\widehat{q},\widehat{p})$, $\widehat{p}%
\rightarrow $ $\widehat{p}^{\prime }(\widehat{q},\widehat{p})$, which
preserves the Dirac bracket $[\widehat{q},\widehat{p}]=i=[\widehat{q}%
^{\prime },\widehat{p}^{\prime }]$.}.

Notably, the inverse transform is given by the same expression with $\mathbf{%
M}$ replaced by the inverse $\mathbf{M}^{-1}$: $\widehat{\mathsf{T}}_{_{\mathbf{M}}}^{-1}=\widehat{\mathsf{T}}_{_{\mathbf{M}^{-1}}}$, thus amounting
to the kernel $\mathsf{K}_{_{\mathbf{M}^{-1}}}(x,x^{\prime })=\mathsf{K}_{_{_{\mathbf{M}}}}^{*}(x^{\prime },x)$.

As said, the Collins integral (\ref{uabcd}) is a real linear CT, the relevant matrix $\mathbf{M}$ signifying in fact the transformation of the paraxial-ray
variables consequent to the propagation through the optical system described
by~$\mathbf{M}$.

The case $B=0$, which optically signif\/ies imaging, yields the geometric
transform
\begin{gather}
\widehat{\mathsf{T}}_{_{\left( _{C\,\,D}^{A\,\,\,0}\right) }}: \ \varphi \in
\mathfrak{L}^2(\mathbb{R})\rightarrow \frac 1{\sqrt{A}}e^{i\frac C{2A}x^2}\varphi
\left(\frac xA\right)\in \mathfrak{L}^2(\mathbb{R}).  \label{geom}
\end{gather}

The extension of real CTs to complex CTs is rather involved. For an accurate
account, we address the reader to the devoted literature \cite{bargmann,moshinsky,quesne,wolfl,wolfr,kramer}. Here, we brief\/ly recall that complex linear CTs amount to
the same integral transform (\ref{can2}), (\ref{can3}), but the kernel
involves the three linearly independent entries of a $2\times 2$
unimodular complex matrix $\mathbf{M}$, belonging to the six-parameter complex
symplectic group $Sp(2,\mathbb{C})\simeq SL(2,\mathbb{C})$; accordingly, the integrability condition requires now that  $\Im{(A/B)}\geq 0$ or $B$ real if $A=0$. In fact, when
extending the concept of canonical transformation from classical to quantum
mechanics, it seemed useful to extend the transformation from the real to
the complex domain. This implies that Hermitian operators can be mapped into
canonically conjugate, but not necessarily Hermitian, operators. Then,
contrary to the real CTs, complex CTs no longer represent unitary mappings
from~$\mathfrak{L}^2(\mathbb{R})$ to~$\mathfrak{L}^2(\mathbb{R})$; they transfer $\mathfrak{L}%
^2(\mathbb{R})$ into the Bargmann--Hilbert space $\mathfrak{F}_{_{\mathbf{M}}}$ of
analytic square-integrable functions over the complex plane~\cite{bargmann},
completed by a suitably def\/ined scalar product in order that the transformed
operators have the appropriate hermiticity properties and reproduce in the
new variable the Schr\"{o}dinger representation as $x$ and $-id/dx$.
Specif\/ically, the complex CT $\widehat{\mathsf{T}}_{_{\mathbf{M}}}$
associated with the transformation matrix $\mathbf{M}\in Sp(2,\mathbb{C)}$
amounts to the transform pair~\cite{wolfl}
\begin{gather}
\widetilde{\varphi }(x)=\big[\widehat{\mathsf{T}}_{_{\mathbf{M}%
}}\varphi \big](x)=\int_{\mathbb{R}}dx^{\prime } \mathsf{K}_{_{_{\mathbf{M}%
}}}(x,x^{\prime })\varphi (x^{\prime }),\nonumber\\
\varphi (x^{\prime })=\int_{\mathbb{C}}d\mu _{_{_{\mathbf{M}%
}}}(x) \mathsf{K}_{_{_{\mathbf{M}}}}^{*}(x,x^{\prime })\widetilde{%
\varphi }(x),\label{can7}
\end{gather}
connecting $\mathfrak{L}^2(\mathbb{R})$ and $\mathfrak{F}_{_{\mathbf{M}}}$, where
\begin{gather*}
\big(\widetilde{\varphi },\widetilde{\psi }\big)_{_{\mathfrak{F}_{_{_{\mathbf{M}}}}}}
=\int_{\mathbb{C}} d\mu _{_{_{\mathbf{M}}}}(x)\widetilde{\varphi }%
^{*}(x)\widetilde{\psi }(x),
\end{gather*}
the measure being
\begin{gather}
d\mu _{_{_{\mathbf{M}}}}(x)=\sqrt{\frac 2{\pi v}}e^{\frac
1{2v}(ux^2-2xx^{*}+u^{*}x^{*2})}\,d\Re{x}\,d\Im{x},\nonumber
\end{gather}
with $u=A^{*}D-B^{*}C\in \mathbb{C}$ and $v=2\Im{(B^{*}A)}\in \mathbb{R}$
\cite{bargmann,moshinsky,quesne,wolfl,wolfr,kramer}.

A basic example of complex linear CT is provided by the Bargmann transform
\cite{bargmann}
\begin{gather}
\big[ \widehat{\mathbb{B}}\varphi \big](x):=\frac 1{\pi ^{1/4}}\int_{-\infty
}^{+\infty }dx^{\prime }e^{-(x^{\prime 2}+x^2-2\sqrt{2}xx^{\prime
})/2}\varphi (x^{\prime })=(2\pi )^{1/4}\big[\widehat{\mathsf{T}}_{_{\mathbf{B}%
}}\varphi \big](x),  \label{bar}
\end{gather}
the relevant canonical transformation, signif\/ied by the Bargmann matrix
\begin{gather*}
\mathbf{B}=\frac 1{\sqrt{2}}
\begin{pmatrix}
1 & -i \\
-i &  1
\end{pmatrix} ,  %\label{bargmannmatrix}
\end{gather*}
essentially turning the coordinate and momentum operators into the harmonic
oscillator raising and lowering operators. Accordingly, the Bargmann
transform relates the Schr\"{o}dinger and Fock representations of quantum
mechanics; it is in fact applied, for instance, to the coherent-state
formulation of quantum optics. Evidently, $\mathfrak{F}_{_{\mathbf{B}}}$ is
the space of holomorphic functions on $\mathbb{C}$ which are square integrable
with respect to a Gaussian measure.

We will specif\/ically deal below with two kinds of complex linear CTs,
associated with the transformation matrices
\begin{gather}
\mathbf{P}(\tau )=
\begin{pmatrix}
1 & -i\tau \\
 0 &  1
\end{pmatrix} ,\qquad \tau >0,  \label{pmatrix}
\end{gather}
and
\begin{gather}
\mathbf{G}(1/w)=
\begin{pmatrix}
1 & 0 \\
i/w & 1
\end{pmatrix},\qquad w>0.   \label{gmatrix}
\end{gather}
The former produces the Poisson (or, Weierstrass--Gauss) transform
\begin{gather*}
\big[ \widehat{\mathsf{T}}_{_{\mathbf{P}}}\varphi \big](x):=\frac 1{\sqrt{2\pi
\tau }}\int_{-\infty }^{+\infty }e^{-\frac 1{2\tau }(x-x^{\prime
})^2}\varphi (x^{\prime })dx^{\prime },  %\label{tp}
\end{gather*}
which may account, for instance, for heat conduction (Section~\ref{section4.1}). It may be
considered as the complex counterpart of the Fresnel transform (\ref{hfi})
accounting for paraxial free-propagation, the relevant transformation matrix
being~(\ref{fsmatrix}). Similarly, it transfers the initial function $%
\varphi (x)$ at $\tau =0$ into the temperature function $\psi (x,\tau )=[%
\widehat{\mathsf{T}}_{_{\mathbf{P}}}\varphi ](x)$ at subsequent~$\tau $s.
Interestingly, since the Poisson transform formally implies the convolution
with a pure Gaussian function, it is optically implementable by a Fourier
transform followed by the propagation through a Gaussian aperture followed
in turn by an inverse Fourier transform. The parameter $1/\tau $ takes the
meaning as the characteristic width of the involved Gaussian function.

As to (\ref{gmatrix}), we see that according to (\ref{geom}) it amounts to a
modulation by the Gaussian $\exp (-x^2/2w)$, which is optically realizable
by propagation through a Gaussian aperture, the parameter $w$ signifying
then the characteristic width of the aperture. Evidently, the real matrix $%
\mathbf{L}(1/f)=\left( _{-1/f}^{\,\,\,\,\,1}\,_{\,\,\,1}^{\,\,0}\right) $,
amounting to the multiplication by the phase factor $\exp (-ix^2/2f)$, is
the real counterpart of (\ref{gmatrix});  it signif\/ies indeed the
propagation through a thin lens.

As their real counterparts, the above matrices obey the semigroup property: $%
\mathbf{P}(\tau _1)\mathbf{P}(\tau _2)=\mathbf{P}(\tau _2)\mathbf{P}(\tau
_1)=\mathbf{P}(\tau _1+\tau _2)$ and $\mathbf{G}(1/w_1)\mathbf{G}(1/w_2)=%
\mathbf{G}(1/w_2)\mathbf{G}(1/w_1)=\mathbf{G}(1/w_1+1/w_2)$, and represent a
dual pair of operators, since $\mathbf{P}(\tau )=\mathbf{F}^{-1}\mathbf{G}%
(\tau )\mathbf{F}$ and similarly $\mathbf{G}(1/w)=\mathbf{F}^{-1}\mathbf{P}%
(1/w)\mathbf{F}$.

An interesting pair of related real/complex CTs is represented by the
Fourier and bilateral Laplace transform, whose features of interest here
will be reviewed below.

\subsubsection{The bilateral Laplace transform}\label{section3.1.1}

The bilateral Laplace transform $\widehat{\mathbb{L}}$ is well known to
transfer an in general complex-valued function def\/ined over $\mathbb{R}$ into a
complex-valued function def\/ined over $\mathbb{C}$ by the integral
\begin{gather}
\big[ \widehat{\mathbb{L}}\varphi \big](x):=\int_{-\infty }^{+\infty }dx^{\prime
}e^{-xx^{\prime }}\varphi (x^{\prime }),  \label{laplacedef}
\end{gather}
\thinspace \thinspace simply referred to as Laplace transform in the
following.

It can be recovered within the formalism of the complex linear CTs. In fact,
according to~(\ref{can7}), the unimodular complex matrix
\begin{gather}
\mathbf{L}=
\begin{pmatrix}
0 & i \\
 i & 0
\end{pmatrix},  \label{laplacematrix}
\end{gather}
yields the integral transform $\widehat{\mathsf{T}}_{\mathbf{L}}$, which is
of\/f of $\widehat{\mathbb{L}}$ by the factor $-i/\sqrt{2\pi }$, being
\begin{gather}
\big[\widehat{\mathsf{T}}_{\mathbf{L}}\varphi \big](x)=\frac 1{i\sqrt{2\pi }%
}\int_{-\infty }^{+\infty }dx^{\prime }e^{-xx^{\prime }}\varphi (x^{\prime
})=\frac 1{i\sqrt{2\pi }}\big[\widehat{\mathbb{L}}\varphi \big](x).
\label{laplacetransform}
\end{gather}

As noted in \cite{wolfl}, since $\mathbf{L}$ does not fullf\/il the
aforestated integrability condition that $B$ be real if $A=0$, the relevant
transform might be meaningless as the inherent integral might be divergent.
Indeed, we know that a critical issue in dealing with Laplace transform is
convergence, since $[\widehat{\mathbb{L}}\varphi ](x)$ generally exists only
for some values of $x$, located in the region of convergence (determined by $\varphi $ and $\Re{x}$). However, we will not dwell here on such a
question, since we will not go through the evaluation of the Laplace
transform of specif\/ic functions, but we will exploit its interpretation as a
linear CT with the associated transformation matrix (\ref{laplacematrix})
and the consequent exponential operator representation in terms of the $su(1,1)$-generators~$\widehat{\mathrm{K}}_{-}$ and~$\widehat{\mathrm{K}}_{+}$. In fact, in analogy with the Fourier transform (see (\ref{fourier})), $\widehat{\mathsf{T}}_{\mathbf{L}}$ can be expressed in the form
\begin{gather}
\widehat{\mathsf{T}}_{\mathbf{L}}=e^{(\pi /2)[\widehat{\mathrm{K}}_{-}-\widehat{\mathrm{K}}_{+}]},  \label{exptl}
\end{gather}
and so interpreted as the non-unitary evolution operator describing the
dynamics of a quantum repulsive oscillator observed at the purely imaginary
``time'' $\tau =i\pi /2$ \cite{wolfpde,wolfef}, whereas, as earlier noted,
the Fourier transform can be understood as the unitary evolution operator
describing the dynamics of the quantum attractive oscillator observed at the
real ``time'' $\tau =\pi /2$. In this connection, it may be worth recalling
that the operators $\widehat{\mathrm{K}}_{-}+\widehat{\mathrm{K}}_{+}\propto
\widehat{\mathrm{P}}^2+\widehat{\mathrm{X}}^2$ and~$\widehat{\mathrm{K}}_{-}-
\widehat{\mathrm{K}}_{+}\propto \widehat{\mathrm{P}}^2-\widehat{\mathrm{X}}%
^2 $ span by exponentiation the elliptic and hyperbolic subgroups of the
symplectic group.

Moreover, paralleling the factorizations (\ref{fact3}) of $\widehat{\mathcal{F}}$, we similarly write
\begin{gather}
\widehat{\mathsf{T}}_{\mathbf{L}}=e^{\widehat{\mathrm{K}}_{-}}e^{-\widehat{%
\mathrm{K}}_{+}}e^{\widehat{\mathrm{K}}_{-}}=e^{-\widehat{\mathrm{K}}_{+}}e^{%
\widehat{\mathrm{K}}_{-}}e^{-\widehat{\mathrm{K}}_{+}},  \label{exptl1}
\end{gather}
since, along with (\ref{dis}), one also has
\begin{gather*}
e^{\beta [\widehat{\mathrm{K}}_{-}-\widehat{\mathrm{K}}_{+}]}=e^{\tan (\beta
/2)\widehat{\mathrm{K}}_{-}}e^{-\sin (\beta )\widehat{\mathrm{K}}%
_{+}}e^{\tan (\beta /2)\widehat{\mathrm{K}}_{-}}\\
\phantom{e^{\beta [\widehat{\mathrm{K}}_{-}-\widehat{\mathrm{K}}_{+}]}}{}
=e^{-\tan (\beta /2)\widehat{\mathrm{K}}_{+}}e^{\sin (\beta )\widehat{\mathrm{K}}_{-}}e^{-\tan (\beta /2)
\widehat{\mathrm{K}}_{+}},\qquad \left| \beta \right| <\pi .
\end{gather*}

As is well known, the Fourier transform admits as eigenfunctions the
Hermite--Gauss functions,
\begin{gather*}
u_n(x)=\frac 1{\sqrt{2^nn!\sqrt{\pi }}}e^{-x^2/2}H_n(x)
\end{gather*}
(that in Section~\ref{section2.3} have been seen to be the source functions $sHG(\xi ,0)$ for
the standard Hermite--Gauss modes), being in fact
\begin{gather*}
\widehat{\mathcal{F}} u_n=(-i)^n u_n.
\end{gather*}

In contrast, the eigenfunctions of $\widehat{\mathsf{T}}_{\mathbf{L}}$
involve the Weber--Hermite functions $D_\nu $, which for integer orders turn
into the $u_n$s: $D_n(\sqrt{2}x)=\sqrt{n!\sqrt{\pi }}u_n(x)$~\cite{magnus}.
The eigenvalue equation for~$\widehat{\mathsf{T}}_{\mathbf{L}}$,
\begin{gather*}
\widehat{\mathsf{T}}_{\mathbf{L}}\Phi _{_\mu }^{^{\pm }}=\lambda _{_{\mathbf{L}}}(\mu )\Phi _{_\mu }^{^{\pm }},
\end{gather*}
is in fact solved by
\begin{gather*}
\Phi _{_\mu }^{^{\pm }}(x)=\frac{\Gamma (-i\mu +1/2)}{2^{3/4}\pi }e^{-i\pi
(i\mu +1/2)/4} D_{i\mu -1/2}\big(\pm e^{i3\pi /4}\sqrt{2}x\big),\qquad \lambda _{_{\mathbf{L}}}(\mu )=e^{\pi \mu /2},
\end{gather*}
for real values of $\mu $ \cite{kalnins1,wolfpde,wolfef}. The functions $%
\Phi _{_\mu }^{^{\pm }}$s provide an eigenfunction basis for $\mathfrak{L}^{2}(\mathbb{R)}$ as well as for the Bargmann--Hilbert space $\mathfrak{F}_{_{\mathbf{L}}} $.

In \cite{torretr} a possible def\/inition of \textit{fractional Laplace
transform} has been proposed by applying to the fractional Fourier transform
matrix $\mathbf{F}^\alpha $ the same similarity transformation which through
the ordinary Fourier transform matrix $\mathbf{F}$ yields $\mathbf{L}$. In
fact, since\footnote{It may be worth noting that it is also $\mathbf{L}=\mathbf{S}(%
e^{i\pi /2})\mathbf{F}=\mathbf{F}\mathbf{S}(e^{-i\pi /2})$, and hence $\widehat{\mathsf{T}}_{%
\mathbf{L}}$ relates to $\widehat{\mathsf{T}}_{\mathbf{F}\text{ }}$with the
relevant integral involving the imaginary axis. As remarked in \cite{wolfl},
in fact, multiplication of the transformation matrix $\mathbf{M}$ on the
left (on the right) by the ``dilation'' matrix $\mathbf{S}(
e^{i\alpha })$ ($\mathbf{S}(e^{-i\alpha })$)
is a tool to obtain transforms involving line integrals along a path tilted
by the phase $\alpha $.}
\[
\mathbf{L}=\mathbf{S}\big(e^{i\pi /4}\big)\mathbf{FS}\big(e^{-i\pi /4}\big),
\]
where $\mathbf{S}$ is the ``dilation'' matrix $\mathbf{S}(e^{i\pi /4})= \binom{\,e^{i\pi /4}\,\,0\,\,}{\,0\,\,\,\,e^{-i\pi /4}}\in SL(2,\mathbb{C}%
)\simeq Sp(2,\mathbb{C})$, one can think of reproducing the above scheme with $%
\mathbf{F}^\alpha $ in place of~$\mathbf{F}$. The resulting matrix
\begin{gather*}
\mathbf{L}^\alpha =\mathbf{S}\big(e^{i\pi /4}\big) \mathbf{F}^\alpha \mathbf{S}
\big(e^{-i\pi /4}\big)=
\begin{pmatrix}
 \cos \phi &  i\sin \phi \\
 i\sin \phi &  \cos \phi
\end{pmatrix},\qquad \phi =\alpha \pi /2,  %\label{flaplacematrix}
\end{gather*}
has been taken in \cite{torretr} as the representative matrix of the
fractional Laplace transform $\widehat{\mathcal{L}}^\alpha $ of order~$\alpha $ by the def\/inition
\begin{gather*}
\widehat{\mathcal{L}} ^\alpha = i^{\alpha /2} \widehat{\mathsf{T}}_{\mathbf{L}^\alpha }.  %\label{flaplaceoperator}
\end{gather*}
Then, echoing the fractional Fourier transform (see~(\ref{falphaoperator2})), $\widehat{\mathcal{L}}^\alpha $ amounts to the integral transform
\begin{gather}
\big[ \widehat{\mathcal{L}}^\alpha \varphi \big](x)=\sqrt{\frac{1-i\cot
\phi }{2\pi i}}\int_{-\infty }^{+\infty }dx^{\prime } e^{\frac 1{2\sin
\phi }(x^{\prime 2}\cos \phi +x^2\cos \phi -2xx^{\prime })}\varphi
(x^{\prime }).  \label{flaplacedef}
\end{gather}
The ordinary transform is recovered with $\alpha =1$, which yields $\widehat{%
\mathcal{L}}^1=i^{1/2}\widehat{\mathsf{T}}_{\mathbf{L}}=(2\pi i)^{-1/2}\widehat{\mathbb{L}}$.

In analogy with $\widehat{\mathcal{F}}^\alpha $, one can prove for $%
\widehat{\mathcal{L}}^\alpha $ the relations
\begin{gather*}
\widehat{\mathcal{L}}^{\alpha +4j}=\widehat{\mathcal{L}}^\alpha ,\qquad j=0,\pm 1,\pm 2,\dots, \\
\widehat{\mathcal{L}}^\alpha =(\widehat{\mathcal{L}})^\alpha,
\end{gather*}
the periodicity with respect to the order $\alpha $ allowing us to limit it
to the interval $(-2,2]$, whereas the power-like relation implies that $\widehat{\mathcal{L}}^\alpha $ admits the same eigenfunctions $\Phi _{_\mu
}^{^{\pm }}$ as $\widehat{\mathcal{L}}$ (and so $\widehat{\mathsf{T}}_{\mathbf{L}}$):
\begin{gather*}
\widehat{\mathcal{L}}^\alpha \Phi _{_\mu }^{^{\pm }}=\lambda _{_{\mathcal{L%
}^\alpha }}(\mu )\Phi _{_\mu }^{^{\pm }},\qquad \lambda _{_{\mathcal{L}^\alpha }}(\mu )=i^{\alpha /2} [\lambda _{_{\mathbf{L}}}(\mu )]^\alpha
=i^{\alpha /2}e^{\pi \alpha \mu /2}.
\end{gather*}

It has recently been noted \cite{sharma} that the transform $\widehat{\mathcal{L}}^\alpha $, as proposed in \cite{torretr} through (\ref{flaplacedef}), does not reproduce the fractional Fourier transform (\ref{falphaoperator2}) when the real part of the transform variable $x^{\prime }=%
\Re{x}^{\prime }+i\Im{x}^{\prime }$ is set to zero. Then, in order
to overcome such a limit, a dif\/ferent parameter matrix, precisely $\mathbf{L}%
^{\prime \alpha }=$ $\binom{i\cos \phi \,\,\,\,\,\,\,\,\,i\sin \phi }{i\sin
\phi \,\,\,\,\,\,\,-i\cos \phi }$, has been suggested in the quoted
reference to def\/ine the fractional Laplace transform through the usual
scheme (\ref{can7}) pertaining to the linear~CTs. We have mentioned such a
result for completeness' sake, but we will not use the proposed matrix $\mathbf{L}^{\prime \alpha }$ in the forthcoming analysis. Actually, also $\mathbf{L}^\alpha $ will marginally be used below.

\subsubsection{Some about the Laplace transform representative matrix}\label{section3.1.2}

The representative matrix of the Laplace transform, be it $\mathbf{L}$ or $\mathbf{L}^\alpha $,
if we are dealing with the fractional transform (\ref{flaplacedef}), belongs to the set of complex unimodular matrices as
\begin{gather}
\mathbf{M}=
\begin{pmatrix}
 A & iB \\
-iC & D
\end{pmatrix},  \label{elle1}
\end{gather}
with $A$, $B$, $C$, $D$ real, and $AD-B$ $C=1$, which indeed form a subgroup
since the product of two matrices of this type yields a matrix of the same
form.

Complex canonical transformations, conveyed by matrices like~(\ref{elle1}),
and their nonunitary representations have a number of interesting
applications in physics. They arise, for instance, in the clustering theory
of nuclei~\cite{kramer} as well as in the study of the accidental degeneracy
\cite{louck} in both a 2D anisotropic oscillator, whose frequencies in the
two directions have a rational ratio, and a 2D isotropic oscillator,
constrained to move in a sector of angle $\pi /q$, $q$ integer.

Also, the matrices (\ref{pmatrix}) and (\ref{gmatrix}), representing optical
processes like the convolution by a~Gaussian function and the propagation
through a Gaussian aperture, are of the type (\ref{elle1}).

Specif\/ically, the matrices (\ref{elle1}) produce the integral transform
\begin{gather}
\big[\widehat{\mathsf{T}}_{_{_{_{\left(
_{-iC\,\,\,D}^{\,\,\,\,A\,\,\,\,\,iB}\right) }}}}\varphi \big](x)=\frac 1{\sqrt{%
-2\pi B}}\int_{-\infty }^{+\infty }e^{\frac 1{2B}(A x^{\prime
 2}+Dx^2-2xx^{\prime })}\varphi (x^{\prime })dx^{\prime }.
\label{gentransl1}
\end{gather}

As the Huygens--Fresnel transform (\ref{uabcd}), it can be recast in a form
similar to~(\ref{vwn}). In fact, (\ref{vwn}) ref\/lects the factorization
\textit{\`{a} la} Wei--Norman of real symplectic matrices~\cite{wei},
\begin{gather*}
\begin{pmatrix}
 A & B \\
C & D
\end{pmatrix} =
\begin{pmatrix}
 1 & 0 \\
C/A & 1
\end{pmatrix}
\begin{pmatrix}
A & 0 \\
0 & 1/A
\end{pmatrix}
\begin{pmatrix}
 1 &  B/A \\
0 & 1
\end{pmatrix},  %\label{elle5}
\end{gather*}
in terms of a dilation matrix sandwiched between two real triangular
matrices, respectively lower-left and upper-right, which optically account
for lensing and free-propagation.

A similar factorization of matrices of the type (\ref{elle1}) is allowed as
\begin{gather}
\begin{pmatrix}
 A & iB \\
-iC & D
\end{pmatrix} =
\begin{pmatrix}
 1  & 0 \\
i/w & 1
\end{pmatrix}
\begin{pmatrix}
A & 0 \\
0 & 1/A
\end{pmatrix}
\begin{pmatrix}
1 & -i\tau \\
0 &  1
\end{pmatrix} ,  \label{ellefact}
\end{gather}
with the real parameters $\tau $ and $w$ given by
\[
\tau =-\frac BA,\qquad \frac 1w=-\frac CA.
\]

Thus, by (\ref{ellefact}) the transform (\ref{gentransl1}) can alternatively
be understood as
\begin{gather}
\big[\widehat{\mathsf{T}}_{_{_{_{\left(
_{-iC\,\,D}^{\,\,\,A\,\,\,iB}\right) }}}}\varphi \big](x)=\frac 1{\sqrt{A}%
}e^{\frac C{2A}x^2}\varphi \left(\frac xA,\tau _0-\frac BA\right),  \label{altern}
\end{gather}
where $\varphi (x,\tau _0-B/A)$ is to be intended as the
``Poisson''-transformed (i.e.\ in a sense, ``evolved'' with respect to the
ef\/fective or f\/ictitious parameter $\tau _0$) form of the function $\varphi
(x)$, namely
\[
\varphi (x,\tau _0+\tau )=\frac 1{\sqrt{2\pi \tau }}\int_{-\infty }^{+\infty
}e^{-\frac 1{2\tau }(x-x^{\prime })^2}\varphi _0(x^{\prime })dx^{\prime }=\big[
\widehat{\mathsf{T}}_{_{\mathbf{P}}}\varphi _0\big](x).
\]
with $\varphi _0(x)=\varphi (x,\tau _0)$.

As for real matrices, the factorization (\ref{ellefact}) holds for $A\neq 0$. If $A=$ $0$, the general expression~(\ref{gentransl1}) yields
\begin{gather}
\big[\widehat{\mathsf{T}}_{_{_{_{\left(
_{-iC\,\,D}^{\,\,\,\,0\,\,\,\,\,iB}\right) }}}}\varphi \big](x)=\frac 1{\sqrt{B}
}e^{\frac D{2B}x^2}\big[\widehat{\mathsf{T}}_{_{\mathbf{L}}}\varphi \big]\left(\frac xB\right),
\label{altern0}
\end{gather}
which involves the Laplace transform (\ref{laplacedef}) of the function.

Of course, the interpretation (\ref{altern}) of (\ref{gentransl1}) can as
well be obtained by directly manipulating the expression of the transform.

\subsection{Radial canonical transforms}\label{section3.2}

Besides the \textit{linear} transforms, \textit{radial} transforms involving
(square-integrable) functions def\/ined on the positive half-line $\mathbb{R}^{+}$
can as well be associated with canonical transformations.

As theorized in \cite{wolfr}, the radial CT $\widehat{\mathsf{T}}_{_{\mathbf{M}}}^{(n,m)}$, generated by the canonical transformation of the $n$-vectors $\widehat{\mathbf{x}}=\left\{ \widehat{x}_j\right\} $ and $\widehat{\mathbf{p}}=\left\{ \widehat{p}_j\right\} $, $j=1,\dots,n$, which, preserving the Dirac
brackets $\left[ \widehat{x}_j,\widehat{p}_k\right] =i\delta _{j,k}$, is
identif\/ied by a $2n\times 2n$ unimodular complex matrix $\mathbf{M}=\left( _{\mathbf{C}}^{\mathbf{A}} \, _{\mathbf{D}}^{\mathbf{B}}\right) $, whose entries
are (complex) multiplies of the $n\times n$ unit matrix such that $AD-BC=1$,
amounts to the transform pair
\begin{gather}
\widetilde{\varphi }(r)=\big[ \widehat{\mathsf{T}}_{_{\mathbf{%
M}}}^{(n,m)}\varphi \big] (r)=\int_0^{+\infty }dr^{\prime } \mathsf{K}
_{_{\mathbf{M}}}^{(n,m)}(r,r^{\prime })r^{\prime  n-1}\varphi (r^{\prime }),\nonumber\\
\varphi (r^{\prime })=\int_{\mathbb{C}^{+}}d\mu _{_{_{\mathbf{M}}} }^{(n,m)}(r) \mathsf{K}_{_{\mathbf{M}}}^{(n,m)}(r,r^{\prime })^{*}\widetilde{\varphi }(r), \label{radial1}
\end{gather}
the kernel being
\begin{gather}
\mathsf{K}_{_{\mathbf{M}}}^{(n,m)}(r,r^{\prime })=\frac{(-i)^{m+n/2}}%
B(rr^{\prime })^{1-n/2}e^{\frac i{2B}(Ar^2+Dr^{\prime  2})}J_{n/2+m-1}\left(\frac{rr^{\prime }}B\right).  \label{radial2}
\end{gather}

By (\ref{radial1}), functions $\varphi (r^{\prime })$ in the Hilbert space $%
\mathcal{L}^2(\mathbb{R}^{+})$ with scalar product
\[
(\varphi ,\psi )_{\mathcal{L}^2(\mathbb{R}^{+})}=\int_0^{+\infty }r^{\prime
 n-1}\varphi ^{*}(r^{\prime })\psi (r^{\prime })dr^{\prime },
\]
are transformed into functions $\widetilde{\varphi }(r)$, belonging to the
space $\mathfrak{F}_{_{\mathbf{M}}}^{(n,m)}$ of analytic functions of the
complex variable $r$ restricted to the region $\mathbb{C}^{+}$, for which $\frac \partial {\partial r^{*}}\widetilde{\varphi }(r)=0$. Moreover, the
space $\mathfrak{F}_{_{\mathbf{M}}}^{(n,m)}$ is equipped with the scalar product
\begin{gather*}
(\widetilde{\varphi },\widetilde{\psi })_{_{\mathfrak{F}_{_{\mathbf{M}%
}}^{(n,m)}}}=\int_{\mathbb{C}^{+}} d\mu _{_{_{\mathbf{M}}}}^{(n,m)}(r)%
\widetilde{\varphi }^{*}(r)\widetilde{\psi }(r),
\end{gather*}
the measure being
\begin{gather*}
d\mu _{_{_{\mathbf{M}}}}^{(n,m)}(r)=\frac 2{\pi v}e^{\frac
1{2v}(ur^2+u^{*}r^{*2})}(rr^{*})^{n/2}K_{n/2+m-1}\left(\frac{rr^{*}}v\right)d\Re{r}\,d\Im{r},
\end{gather*}
where $K_\nu $ denotes the MacDonald function of order $\nu $ \cite{magnus},
and, as before, $u=A^{*}D-B^{*}C\in \mathbb{C}$ and $ v=2\Im{(B^{*}A)}\in
\mathbb{R}$ \cite{wolfl,wolfr}. It ensures that the transformed operators have
the appropriate hermiticity properties and reproduce in the new variable the
Schr\"{o}dinger representation of the operators $\widehat{\mathbf{x}}^2=r^2$, $\widehat{\mathbf{x}}\cdot \widehat{\mathbf{p}}=-ir\frac \partial
{\partial r}$ and $\widehat{\mathbf{p}}^2=-(\frac{\partial ^2}{\partial r^2}+
\frac{n-1}r\frac \partial {\partial r}+\frac \lambda {r^2})$, with the real
parameter $\lambda =-m(m+n-2)$, $m=0,1,2,\dots$, addressing the eigenvalue of
the angular momentum $\widehat{L}^2=\frac 12\sum \widehat{L}_{jk}\widehat{L}%
_{jk}$, $\widehat{L}_{jk}=\widehat{x}_j\widehat{p}_k-\widehat{x}_k\widehat{p}%
_j$, which is invariant under the unimodular transformation~$\mathbf{M}$~\cite{wolfr}.

In particular, with $n=2$ and $\mathbf{M}$ resorting to the entries of the
Fourier transform matrix~$\mathbf{F}$, one basically obtains the Hankel
transform~(\ref{hankel}), whereas the fractional Fourier transform matrix~$\mathbf{F}^\alpha $ yields through~(\ref{radial2}) the fractional Hankel
transform~(\ref{hfr}).

The Barut--Girardello transform $\widehat{\mathbb{G}}_{n,m}$ represents an
interesting example of complex radial~CT. Resorting to the Bargmann matrix $\mathbf{B}$, it signif\/ies \cite{barut}
\begin{gather}
\big[\widehat{\mathbb{G}}_{n,m}\varphi \big](r)=\sqrt{2}\int_0^{+\infty
}dr^{\prime } (rr^{\prime })^{1-n/2}e^{-\frac 12(r^2+r^{\prime
 2})}I_{n/2+m-1}(\sqrt{2}rr^{\prime })\varphi (r^{\prime
})\nonumber\\
\phantom{\big[\widehat{\mathbb{G}}_{n,m}\varphi \big](r)}{}
=i^{n/2+m-1}\big[ \widehat{\mathsf{T}}_{_{\mathbf{B}}}^{(n,m)}\varphi
\big] (r),  \label{bgt}
\end{gather}
$I_\nu $ denoting the modif\/ied Bessel function of the f\/irst kind of order $%
\nu $: $J_\nu (ix)=i^\nu I_\nu (x)$~\cite{magnus}. It was introduced in
developing the formalism for coherent states associated with Lie algebras of
non-compact groups, like, in particular, the semisimple Lie algebra $%
so(2,1)\simeq sl(2,\mathbb{R})$. As a~direct generalization of the notion of
coherent states associated with the Heisenberg algebra, such \textit{generalized coherent states} were introduced as eigenstates of the lowering
operator of the aforesaid algebra in the relative discrete representations $%
D^{\pm }(j)$, $j=-1/2,-1,-3/2,\dots$~\cite{barut}.

\subsection{Essentials of Hankel-type transforms}\label{section3.3}

The f\/irst and second Hankel-type transforms of Bessel order $\nu \geq -1/2$,
depending on an arbitrary real parameter $\nu ^{\prime }$, are respectively
def\/ined by \cite{linares,malgonde2}
\begin{gather}
\big[ \widehat{\mathcal{H}}_{_{1,\nu ,\nu ^{\prime
}}}f\big] (y)\equiv \widetilde{f}_{_{1,\nu ,\nu ^{\prime }}}(y):=y^{1+2\nu
^{\prime }}\int_0^{+\infty }(xy)^{-\nu ^{\prime }}J_\nu (xy)f(x)dx,\nonumber\\
\big[ \widehat{\mathcal{H}}_{_{2,\nu ,\nu ^{\prime }}}f\big]
(y)\equiv \widetilde{f}_{_{2,\nu ,\nu ^{\prime }}}(y):=\int_0^{+\infty
}x^{1+2\nu ^{\prime }}(xy)^{-\nu ^{\prime }}J_\nu (xy)f(x)dx,\label{defh}
\end{gather}
the function $f(x)$ being assumed to belong to the space $\mathcal{L}^2(\mathbb{%
R}^{+})$.

It can be seen that they relate to each other according to
\begin{gather*}
\widehat{\mathcal{H}}_{_{1,\nu ,\nu ^{\prime }}}^{\dagger }=\widehat{%
\mathcal{H}}_{_{2,\nu ,\nu ^{\prime }}},\qquad \widehat{\mathcal{H}}_{_{2,\nu
,\nu ^{\prime }}}^{\dagger }=\widehat{\mathcal{H}}_{_{1,\nu ,\nu ^{\prime
}}}.
\end{gather*}
Also, as a consequence of the well-known orthogonality relation of the
Bessel functions, both transforms are self-reciprocal:
\begin{gather*}
\widehat{\mathcal{H}}_{_{1,\nu ,\nu ^{\prime }}}^{-1}=\widehat{\mathcal{H}}%
_{_{1,\nu ,\nu ^{\prime }}},\qquad \widehat{\mathcal{H}}_{_{2,\nu
,\nu ^{\prime }}}^{-1}=\widehat{\mathcal{H}}_{_{2,\nu ,\nu ^{\prime }}},
\end{gather*}
and separately obey the Parseval relations \cite{malgonde3}
\begin{gather*}
\int_0^{+\infty }x^{-1-2\nu ^{\prime
}}f^{*}(x)g(x)dx=\int_0^{+\infty }x^{-1-2\nu ^{\prime }}\widetilde{f}%
_{_{1,\nu ,\nu ^{\prime }}}^{*}(x)\widetilde{g}_{_{1,\nu ,\nu ^{\prime
}}}(x)dx, \\
\int_0^{+\infty }x^{1+2\nu ^{\prime
}}f^{*}(x)g(x)dx=\int_0^{+\infty }x^{1+2\mu }\widetilde{f}_{_{2,\nu ,\nu
^{\prime }}}^{*}(x)\widetilde{g}_{_{2,\nu ,\nu ^{\prime }}}(x)dx,
\end{gather*}
along with the mixed relation
\begin{gather*}
\int_0^{+\infty }f^{*}(x)g(x)dx=\int_0^{+\infty }\widetilde{f}_{_{1,\nu
,\nu ^{\prime }}}^{*}(x)\widetilde{g}_{_{2,\nu ,\nu ^{\prime }}}(x)dx.
\end{gather*}

In \cite{torreh} the corresponding fractional order transforms have been
introduced.

Notably, one can prove the identities \cite{linares,malgonde2}
\begin{gather}
\big[ \widehat{\mathcal{H}}_{_{1,\nu ,\nu ^{\prime }}}%
\widehat{B}_{_{\nu ,\nu ^{\prime }}}^{\dagger }f\big] (y)=-y^2\big[
\widehat{\mathcal{H}}_{_{1,\nu ,\nu ^{\prime }}}f\big] (y),\qquad
\big[ \widehat{\mathcal{H}}_{_{2,\nu ,\nu ^{\prime }}}\widehat{B%
}_{_{\nu ,\nu ^{\prime }}}f\big] (y)=-y^2\big[ \widehat{\mathcal{H}}%
_{_{2,\nu ,\nu ^{\prime }}}f\big] (y),\label{eveq1}
\end{gather}
involving the Bessel-type dif\/ferential operator $\widehat{B}_{_{\nu ,\nu
^{\prime }}}$ and its adjoint $\widehat{B}_{_{\nu ,\nu ^{\prime }}}^{\dagger
}$:
\begin{gather*}
\widehat{B}_{_{\nu ,\nu ^{\prime }}}:=\frac{d^2}{dx^2}+(1+2\nu
^{\prime })\frac 1x\frac d{dx}+\big(\nu ^{\prime 2}-\nu ^2\big)\frac
1{x^2}=x^{-\nu ^{\prime }-\nu -1}\frac \partial {\partial x}x^{2\nu +1}\frac
\partial {\partial x}x^{\nu ^{\prime }-\nu },\nonumber\\
\widehat{B}_{_{\nu ,\nu ^{\prime }}}^{\dagger }:=\frac{d^2}{dx^2}
-(1+2\nu ^{\prime })\frac 1x\frac d{dx}+\big[(\nu ^{\prime }+1)^2-\nu ^2\big]\frac
1{x^2}=x^{\nu ^{\prime }-\nu }\frac \partial {\partial x}x^{2\nu +1}\frac
\partial {\partial x}x^{-\nu ^{\prime }-\nu -1}.%\label{bop}
\end{gather*}
Relations (\ref{eveq1}) can be used, for instance, to write the solution of
the evolutionary equation \cite{linares,malgonde2}
\begin{gather}
\kappa \frac \partial {\partial \tau }h(x,\tau )=\widehat{B}_{_{\nu ,\nu
^{\prime }}}^{\dagger }h(x,\tau ),  \label{evoleq}
\end{gather}
satisfying the initial condition $h(x,0)=f(x)$, in the transform conjugate $%
y $-space as $\widetilde{h}_{_{1,\nu ,\nu ^{\prime }}}(y,\tau )=e^{-(\tau
/\kappa )y^2}\widetilde{f}_{_{1,\nu ,\nu ^{\prime }}}(y)$ for arbitrary
values of $\kappa $. Then, transforming back to the $x$-space, we obtain{\samepage
\begin{gather*}
h(x,\tau )=\frac \kappa {2\tau }x^{1+2\nu ^{\prime }}\int_0^{+\infty
}(xy)^{-\nu ^{\prime }}e^{-\frac \kappa {4\tau }(x^2+y^2)}I_\nu \big(\tfrac
\kappa {2\tau }xy\big)f(y)dy,   %\label{solx}
\end{gather*}
under the condition that $\left| \arg (\tau /\kappa )\right| <$ $\pi /4$,
which for both $\tau $ and $\kappa $ real turns into $\tau /\kappa >0$.}

As noted in \cite{torreh}, on account of the symbolic solution $h(x,\tau
)=\exp [(\tau /\kappa )\widehat{B}_{_{\nu ,\nu ^{\prime }}}^{\dagger
}] f(x) $ of (\ref{evoleq}), the above yields an explicit functional
representation of the exponential operator $\exp [\beta \widehat{B}_{_{\nu
,\nu ^{\prime }}}^{\dagger }]$:
\begin{gather}
\big[ e^{\beta \widehat{B}_{_{\nu ,\nu ^{\prime }}}^{\dagger }}f\big]
(x)=\frac 1{2\beta }x^{1+2\nu ^{\prime }}\int_0^{+\infty }(xy)^{-\nu
^{\prime }}e^{-\frac 1{4\beta }(x^2+y^2)}I_\nu \big(\tfrac{xy}{2\beta }\big)f(y)dy.
\label{expop}
\end{gather}
For $\beta =i/2$ it eventually conveys a representation of the transform $%
\widehat{\mathcal{H}}_{_{1,\nu ,\nu ^{\prime }}}$ as a symmetric product of
exponentials of the $su(1,1)$ algebra generators, similar to (\ref{facthankel}) for the Hankel transforms:
\begin{gather}
\widehat{\mathcal{H}}_{_{1,\nu ,\nu ^{\prime }}}=i^{\nu +1}e^{-\frac
i2x^2}e^{\frac i2\widehat{B}_{_{\nu ,\nu ^{\prime }}}^{\dagger }}e^{-\frac
i2x^2}=i^{\nu +1}e^{-i\frac \pi 2[\widehat{\mathrm{K}}_{-}^{(1)}+\widehat{%
\mathrm{K}}_{+}^{(1)}]}.  \label{hsus}
\end{gather}
This is not surprising since the operators
\begin{gather}
\widehat{\mathrm{K}}_{+}^{(1)}:=\tfrac 12x^2,\qquad \widehat{\mathrm{K}}%
_{-}^{(1)}:=-\tfrac 12\widehat{B}_{_{\nu ,\nu ^{\prime }}}^{\dagger
},\qquad \widehat{\mathrm{K}}_3^{(1)}:=-\tfrac i2\left(x\frac d{dx}-\nu ^{\prime
}\right),  \label{su1}
\end{gather}
constitute a non self--adjoint one-variable realization of the $su(1,1)$
algebra generators obeying, as it should be, the inherent commutation
relations shown in~(\ref{su11}).

The adjoint operators
\begin{gather}
\widehat{\mathrm{K}}_{+}^{(2)}:=\tfrac 12x^2,\qquad \widehat{\mathrm{K}}%
_{-}^{(2)}:=-\tfrac 12\widehat{B}_{_{\nu ,\nu ^{\prime }}},\qquad \widehat{%
\mathrm{K}}_3^{(2)}:=-\tfrac i2\left(x\frac d{dx}+\nu ^{\prime }+1\right)  \label{su2}
\end{gather}
pertain to the second Hankel-type transform $\widehat{\mathcal{H}}_{_{2,\nu
,\nu ^{\prime }}}$, which is in fact amenable for a factored form
representation analogous to (\ref{hsus}) in terms of them~\cite{torreh}.

With $\nu ^{\prime }=-1$ and $\nu =m$ the operators~(\ref{su1}) turn into (\ref{su1c}), pertaining to the Hankel transform $\widehat{\mathcal{H}}_m$ of
order~$m$, considered in Section~\ref{section2.2} in connection with the propagation problem
in circular cylindrical coordinates; indeed, $\widehat{\mathcal{H}}_{_{1,\nu
,\nu ^{\prime }}}\rightarrow $ $\widehat{\mathcal{H}}_m$ for the identif\/ied
values of~$\nu $ and~$\nu ^{\prime }$.

In general, as noted in \cite{torreh}, for suitable values of $\nu ^{\prime
} $ and $\nu $ the transforms $\widehat{\mathcal{H}}_{_{1,\nu ,\nu ^{\prime
}}} $ and $\widehat{\mathcal{H}}_{_{2,\nu ,\nu ^{\prime }}}$ can be framed
within the formalism of the radial CTs, as theorized in \cite{wolfr} and
shortly reviewed in Section~\ref{section3.2}. In fact, if the $2n\times 2n$ transformation
matrix resorts to the entries of the Fourier matrix~$\mathbf{F}$, the kernel~(\ref{radial2}) becomes
\begin{gather*}
\mathsf{K}_{_{\mathbf{F}}}^{(n,m)}(r,r^{\prime })=(-i)^{m+n/2}(rr^{\prime
})^{1-n/2}J_{n/2+m-1}(rr^{\prime }),  %\label{radialft}
\end{gather*}
the resulting transform being seen as the radial part of a $n$-dimensional
linear CT, specif\/ically representing a $\pi /2$-rotation for each pair of
the canonically conjugate operators in the relevant $n$-component position
and momentum operator vectors. The kernel $\mathsf{K}_{_{\mathbf{F}%
}}^{(n,m)} $ resembles that of the transforms (\ref{defh}) for suitable
settings of the parameters $\nu $ and $\nu ^{\prime }$ in terms of the
inherent dimension $n$ and eigenvalue $\lambda $. In fact, one respectively
has $n=-2\nu ^{\prime }$ and $n=2(1+\nu ^{\prime })$, which so are
meaningful for~$\nu ^{\prime }$ integer or half-integer (negative or
positive), whereas correspondingly $m=\nu +\nu ^{\prime }+1$ and $m=\nu -\nu
^{\prime }$, thus yielding $\lambda =(\nu ^{\prime }+1)^2-\nu ^2$ and $%
\lambda =\nu ^{\prime 2}-\nu ^2$.

Disregarding the specif\/ic link of the canonical transforms to the canonical
transformations as brief\/ly reviewed above, we rewrite the integral
transforms (\ref{defh}) through the formal relations
\begin{gather}
\big[ \widehat{\mathcal{H}}_{_{1,\nu ,\nu ^{\prime
}}}\varphi \big] (r)=i^{1+\nu }\int_0^\infty \mathsf{K}_{_{\mathbf{F}%
}}^{(-2\nu ^{\prime },\nu +\nu ^{\prime }+1)}(r,r^{\prime })r^{\prime
 -1-2\nu ^{\prime }}\varphi (r^{\prime })dr^{\prime },\nonumber\\
\big[ \widehat{\mathcal{H}}_{_{2,\nu ,\nu ^{\prime }}}\varphi
\big] (r)=i^{1+\nu }\int_0^\infty \mathsf{K}_{_{\mathbf{F}}}^{(2(1+\nu
^{\prime }),\nu -\nu ^{\prime })}(r,r^{\prime })r^{\prime 1+2\nu
^{\prime }}\varphi (r^{\prime })dr^{\prime },\label{defhct}
\end{gather}
which, contrary to the general account (\ref{radial1}), should be intended
as transforming functions of a real variable into functions of a real
variable.

In virtue of the analysis presented in Section~\ref{section2}, one may guess that the above
might be of relevance in connection with the Appell transformation for a
``radial PWE'' in a higher-dimension domain.

In Section~\ref{section4.3} we will prove that the ``Laplace-like'' versions of (\ref{defhct}) relate to the Appell transformation for the radial HE, which is seen to
connect temperature functions evolving from functions linked through the
aforesaid transforms.

\section{Caloric Appell transformation:\\ Laplace and Hankel-type transforms}\label{section4}

In virtue of the well-known analogy between the 1D HE and the 2D linear PWE,
we will properly reproduce within the context of the 1D HE the
interpretation of the Appell transformation, as elaborated in Section~\ref{section2.1.2} in
connection with the optical propagation. We will f\/ind that the well-known
\textit{caloric} Appell transformation connects temperature functions
resulting from initial conditions linked by a Laplace-like similarity
transformation. Thus, as the \textit{optical} Appell transformation is
understood to manifest the action of the evolving-location Fourier transform
operator on wavefunctions, likewise the \textit{caloric} Appell
transformation will be seen to result from the action of the evolving-time
Laplace-like transform operator on temperature functions. A~similar relation
will then be shown to hold also for the Appell transformation for the radial
HE, accordingly resorting to suitably def\/ined Hankel-type transforms,
explicable as ``radial-Laplace''-type transforms.

\subsection{1D heat equation}\label{section4.1}

The 1D~HE\footnote{Actually, the standard form of the 1D~HE is \cite{widderb}
\[
\frac \partial {\partial t}u(x,t)=\frac{\partial ^2}{\partial x^2}u(x,t).
\]
Here, in order to favor the comparison with the 2D PWE, we will work with (%
\ref{heat}), which amounts to the trivial variable scaling: $t\rightarrow
t/2 $ between the ``time'' in the standard form and that in (\ref{heat}).}
\begin{gather}
2\frac \partial {\partial t}u(x,t)=\frac{\partial ^2}{\partial x^2}u(x,t)
\label{heat}
\end{gather}
has been the object of extensive studies; few titles from the wide
bibliography are in \cite{appell,widderb,rosenbloom,widder1,widder2,leutwiler,shimomura}. See also \cite{miller,wolfb,olver}. The \textit{temperature }or \textit{caloric function} $u(x,t)$
is usually assumed to be of class $C^1$ with respect to $t$ and of class $%
C^2 $ with respect to $x$ across the region $\mathcal{D}$ in the $x,t$-plane
where (\ref{heat}) holds~\cite{widderb}. In view of the analysis we will
perform, $u(x,t)$ will be here supposed to be of class $C^\infty $ with
respect to the variables $x$, $t$, which will be considered as real unitless
variables.

As is well known, solutions to (\ref{heat}) can be obtained as the Poisson
transform of the initial conditions, that is \cite{widderb}
\begin{gather}
  f(x,t) =\big[\widehat{\mathcal{U}}_{_{\mathsf{HE}}}(t)f_0\big](x):=\frac 1{\sqrt{%
2\pi t}}\int_{-\infty }^\infty e^{-\frac{(x-y)^2}{2t}}f_0(y)dy,  \label{genh}
\end{gather}
with $f_0(x)=f(x,0)$, under the general statement that $f_0(x)$ be Lebesgue
integrable in every f\/inite interval. The integral kernel reproduces the
fundamental solution of (\ref{heat}):
\begin{gather}
S(x,t):=\frac 1{\sqrt{2\pi t}}e^{-\frac{x^2}{2t}},\qquad t>0.
\nonumber
\end{gather}

Resorting to the review in Section~\ref{section3.1}, we see that $\widehat{\mathcal{U}}_{_{%
\mathsf{HE}}}=\widehat{\mathsf{T}}_{_{\mathbf{P}}}$, with the replacement \mbox{$\tau \rightarrow t$} in the relevant canonical transformation matrix $\mathbf{%
P}$. Evidently, the Poisson transform (\ref{genh}) corresponds to the
Fresnel transform (\ref{hfi}) by the variable interchange $t\leftrightarrow
i\zeta $, which is the basis of the analogy between the 1D~HE and the 2D~PWE. It conveys an integral transform representation for the ``evolution''
operator $\widehat{\mathcal{U}}_{_{\mathsf{HE}}}(t)$ associated with (\ref
{heat}), which can as well be given the exponential operator representation
\begin{gather}
 \widehat{\mathcal{U}}_{_{\mathsf{HE}}}(t):=e^{\frac t2\frac{d^2}{dx^2}%
}=e^{-t\widehat{\mathrm{K}}_{-}},  \label{eoheat}
\end{gather}
in analogy with that of the optical propagator $\widehat{\mathcal{U}}_{_{%
\mathsf{PWE}}}(\zeta )$. As in that case, we consider the two
representations as fully equivalent, disregarding indeed the question of
specifying the domain of functions over which such an equivalence
ef\/fectively holds~\cite{widderb}.

We have expressed $\widehat{\mathcal{U}}_{_{\mathsf{HE}}}(t)$ in terms of
the operator $\widehat{\mathrm{K}}_{-}$, thus signalizing that in connection
with the 1D~HE we will resort to the same basis operators (\ref{su11}) as
for the 2D~PWE, although linear combinations of them by in general complex
coef\/f\/icients will more properly be considered. This will amount to deal with
complex canonical transforms. In fact, on discussing the \textit{caloric}
Appell transformation we will be led to consider the Laplace transform
represented by the complex symplectic matrix $\mathbf{L}$ (Section~\ref{section3.1.1}),
whereas the afore-discussed \textit{optical} Appell transformation has led
to the Fourier transform, which in turn resorts to the real symplectic
matrix $\mathbf{F}$ (Section~\ref{section2.1.2}).

\subsection{Some about the heat polynomials and the associated functions}\label{section4.2}

The heat polynomials $v_n(x,t)$ are def\/ined as the polynomial solutions of (\ref{heat}), which realize the power series expansion of the simple
exponential solution~\cite{widderb}, i.e.
\begin{gather}
e^{\chi x+\chi ^2t/2}=\sum_{n=0}^\infty \frac{\chi ^n}{n!}v_n(x,t),
\label{genfun}
\end{gather}
with $\chi $ arbitrary parameter. They explicitly write as
\begin{gather}
  v_n(x,t) := n!\sum_{j=0}^{[n/2]}\frac{t^j}{2^jj!(n-2j)!} x^{n-2j},
\nonumber
\end{gather}
which reveals their connection with the Hermite polynomials as \mbox{$%
v_n(x,t)\!=\!(-t/2)^{n/2}H_n(x/\sqrt{-2t})$.}

Interestingly, the $v_n$s solve (\ref{heat}) subject to monomial initial
distributions $v_n(x,0) =x^n$, being~\cite{widderb}
\begin{gather}
  v_n(x,t) =\frac 1{\sqrt{2\pi t}}\int_{-\infty }^\infty e^{-\frac{(x-y)^2}{%
2t}}y^ndy,\qquad 0<t<\infty .  \label{hpint}
\end{gather}

In addition, they obey the second-order dif\/ferential equations
\begin{gather}
\left(t\frac{d^2}{dx^2}+x\frac d{dx}-n\right)v_n(x,t)=0,  \label{eqv}
\end{gather}
which determine the $x$-dependence of the $v_n$s, whose complete expressions
follow then from the inherent initial conditions $v_n(x,0)=x^n$.

The associated functions $w_n(x,t)$, which are solutions of the HE as well,
are obtained from the $v_n$s by the \textit{caloric} Appell transformation
\cite{appell,widderb}:
\begin{gather}
w_n(x,t):=S(x,t)v_n\left(\frac xt,-\frac 1t\right)=t^nS(x,t)v_n(x,-t).  \label{ahp}
\end{gather}
They admit the generating function relation
\[
\frac 1{\sqrt{2\pi t}}e^{-\frac{(x-\chi )^2}{2t}}=\sum_{n=0}^\infty \frac{\chi ^n}{n!}w_n(x,t)=S(x-\chi ,t),
\]
which parallels (\ref{genfun}), as well as, in analogy with (\ref{hpint}),
the integral representation \cite{widderb}
\[
w_n(x,t)=\frac 1{2\pi }\int_{-\infty }^\infty e^{-\frac t2y^2+ixy}(-iy)^ndy.
\]

Finally, the $w_n$s obey the dif\/ferential equation
\begin{gather}
\left(t\frac{d^2}{dx^2}+x\frac d{dx}+n+1\right)w_n(x,t)=0.  \label{eqw}
\end{gather}

We address the reader to the bibliography \cite{widderb,rosenbloom,widder1,widder2} for
a more complete account of the topic.

In \cite{torreat} the optical analogs of the heat polynomials and the
associated functions have been deduced, which have suggested the
interpretation of the \textit{optical} Appell transformation in terms of
Fourier-related source functions, as illustrated in Section~\ref{section2.1.2}.

\subsection{Laplace transform and Appell transformation for the 1D heat
equation}\label{section4.3}

As earlier remarked, the suitable Lie-algebra context should be now that of
the six-parameter algebra $sl(2,\Bbb{C})$; accordingly, taking account also
of the Weyl algebra, we should deal with the semigroup $WSL(2,\Bbb{C})$.
However, in our analysis we will make use of CTs of the type~(\ref{gentransl1}) (associated with matrices of the form (\ref{elle1})). Hence,
we can limit ourselves to deal with $SL(2,\Bbb{C})$, even better with the
subgroup $\subset SL(2,\Bbb{C})$ formed by the matrices (\ref{elle1}), that
amounts to a subalgebra $\subset sl(2,\Bbb{C})$, spanned by the basis
operators $\big\{ \widehat{\mathrm{K}}_{+},\widehat{\mathrm{K}}_3,\widehat{%
\mathrm{K}}_{-}\big\} $ through linear combinations with coef\/f\/icients
respectively of the type $\{i\alpha ,\gamma ,i\beta \}$ with $\alpha $, $\beta $, $\gamma $ real.

We consider for every operator $\widehat{\mathrm{K}}$ in the aforesaid
subalgebra $\subset sl(2,\Bbb{C})$ the relevant ``dual'' operator $\widehat{%
\widetilde{\mathrm{K}}}=\widehat{\mathsf{T}}_{\mathbf{L}}\widehat{\mathrm{K}}%
\widehat{\mathsf{T}}_{\mathbf{L}}^{-1}$, obtained from the former by a
similarity transformation involving the Laplace-like operator $\widehat{%
\mathsf{T}}_{\mathbf{L}}{}$. Due to (\ref{exptl}) or (\ref{exptl1}), $%
\widehat{\widetilde{\mathrm{K}}}$ belongs to the same subalgebra as $%
\widehat{\mathrm{K}}$.

By retracing the eigenstate-based procedure \cite{kalnins1,miller} applied
to the 2D PWE in Section~\ref{section2.1.2}, we consider the eigenvalue problems for $%
\widehat{\mathrm{K}}$, as represented by the equivalent eigenvalue equations
\begin{gather}
\widehat{\mathrm{K}}f_\lambda (x)=\lambda f_\lambda (x),  \label{kc}
\end{gather}
or
\begin{gather}
\widehat{\mathrm{K}}(t)v_\lambda (x,t)=\lambda v_\lambda (x,t),  \label{ktc}
\end{gather}
and for $\widehat{\widetilde{\mathrm{K}}}$, as similarly described by
\begin{gather}
\widehat{\widetilde{\mathrm{K}}}g_\lambda (x)=\lambda g_\lambda (x),
\label{ktflc}
\end{gather}
or
\begin{gather}
\widehat{\widetilde{\mathrm{K}}}(t)w_\lambda (x,t)=\lambda w_\lambda (x,t).
\label{ktelc}
\end{gather}
Here, the evolving-time operators explicitly mean
\[
 \widehat{\mathrm{K}}(t):=\widehat{\mathcal{U}}_{_{\mathsf{HE}}}(t)%
\widehat{\mathrm{K}} \widehat{\mathcal{U}}_{_{\mathsf{HE}}}^{-1}(t)=e^{-t%
\widehat{\mathrm{K}}_{-}}\widehat{\mathrm{K}}e^{t\widehat{\mathrm{K}}_{-}}.
\]

Just as for the PWE, the above equations signify that we are looking for
temperature functions $v(x,t)$ and $w(x,t)$ corresponding to initial
conditions linked by the integral transform (\ref{laplacetransform}), i.e.\ $v(x,0)=f(x)$ and $w(x,0)=g(x)=[\widehat{\mathsf{T}}_{\mathbf{L}}f](x)$, with
in particular $f(x)$ being identif\/ied as an eigenstate of a given operator
in the subalgebra of concern. Hence, as $v_\lambda (x,t)=e^{-t\widehat{%
\mathrm{K}}_{-}}f_\lambda (x)$, the $w_\lambda $s are in turn obtained from
the $v_\lambda $s by the ``instantaneous'' transformation
\begin{gather*}
w_\lambda (x,t)=e^{-t\widehat{\mathrm{K}}_{-}}g_\lambda (x)=e^{-t\widehat{%
\mathrm{K}}_{-}}\widehat{\mathsf{T}}_{\mathbf{L}}e^{t\widehat{\mathrm{K}}%
_{-}}v_\lambda (x,t)=\widehat{\mathcal{A}}v_\lambda (x,t).  %\label{wvc}
\end{gather*}
The inherent symmetry operator $\widehat{\mathcal{A}}$, identif\/ied by the
same symbol as the \textit{optical} Appell operator,
\begin{gather*}
\widehat{\mathcal{A}}:=e^{-t\widehat{\mathrm{K}}_{-}}\widehat{\mathsf{T}}_{%
\mathbf{L}}e^{t\widehat{\mathrm{K}}_{-}}=\widehat{\mathsf{T}}_{\mathbf{L}%
}(t),  %\label{appellc}
\end{gather*}
signif\/ies now the evolving-time Laplace operator. Its explicit dependence on
$t$ will be omitted.

Contrary to the \textit{optical} operator, the \textit{caloric} \textit{%
Appell operator} individualizes the complex canonical transformation matrix
\begin{gather*}
\mathbf{M}_{\widehat{\mathcal{A}}}:=
\begin{pmatrix}
1 & -it \\
0 & 1
\end{pmatrix}
\begin{pmatrix}
0 & i \\
i & 0
\end{pmatrix}
\begin{pmatrix}
1 & it \\
0 & 1
\end{pmatrix} =
\begin{pmatrix}
t & i(1+t^2) \\
i & -t
\end{pmatrix} .  %\label{mappellc}
\end{gather*}
Being it of the type~(\ref{elle1}), we can apply the recipe (\ref{altern})
with $\tau _0=t$, thus inferring the asso\-cia\-ted functions~$w_\lambda $s from
the $v_\lambda $s in accord to the Appell rule~(\ref{ahp})  (apart from the
factor~$1/\sqrt{2\pi }$):
\begin{gather*}
w_\lambda (x,t):=\widehat{\mathcal{A}}v_\lambda (x,t)=\frac 1{\sqrt{t}}e^{-\frac{x^2}{2t}}v_\lambda \left(\frac xt,-\frac 1t\right).
%\label{appell1c}
\end{gather*}
By (\ref{altern0}) one recovers the primary relation between the ``source''
functions, i.e.\ $g(x)=[\widehat{\mathsf{T}}_{\mathbf{L}}f](x)$.

Notably, disregarding the formal complexity of the inverse transform and
na\"{\i}vely using the inverse matrix $\mathbf{M}_{\widehat{\mathcal{A}}%
^{-1}}=\binom{-t\,\,\,\,\,-i(1+t^2)}{-i\,\,\,\,\,\,\,\,\,\,\,\,\,\,t\,\,\,\,%
\,\,\,\,}$,  one obtains the correct rule to recover the~$v$s from the~$w$s.

As a basic example, we may consider the heat polynomials and the associated
functions. As signalized by~(\ref{hpint}), the former are caloric functions
arising from the dif\/fusion of the mono\-mials~$x^n$s.  Therefore, the
eigenvalue equations~(\ref{kc}) and~(\ref{ktc}) must be specialized for the
scale transform generator~$\widehat{\mathrm{K}}_3$:
\[
\widehat{\mathrm{K}}_3x^n=\lambda x^n,
\]
with $\lambda =-i(n+1/2)$, $n=0,1,2,\dots$, and for the relevant
evolving-time operator $\widehat{\mathrm{K}}_3(t)$:
\[
\widehat{\mathrm{K}}_3(t)v_n(x,t)=\big[\widehat{\mathrm{K}}_3+it\widehat{\mathrm{K}}_{-}\big]v_n(x,t)=\lambda v_n(x,t),
\]
eventually resulting into (\ref{eqv}), obeyed by the $v_n$s.

As to the dual operator $\widehat{\widetilde{\mathrm{K}}}_3=\widehat{\mathsf{%
T}}_{\mathbf{L}}\widehat{\mathrm{K}}_3\widehat{\mathsf{T}}_{\mathbf{L}%
}^{-1}=-$ $\widehat{\mathrm{K}}_3$, the eigenvalue equation~(\ref{ktflc})
writes as
\[
\widehat{\mathrm{K}}_3g_\lambda =-\lambda g_\lambda ,
\]
whereas (\ref{ktelc}) for the evolving-time operator $\widehat{\widetilde{%
\mathrm{K}}}_3(t)=-\widehat{\mathrm{K}}_3-it\widehat{\mathrm{K}}_{-}$
specializes as
\[
\big[ \widehat{\mathrm{K}}_3+it\widehat{\mathrm{K}}_{-}\big]w_n(x,t)=\frac
i2\left(n+\frac 12\right)w_n(x,t),
\]
which, as expected, yields (\ref{eqw}) obeyed by the associated functions~$w_n$.

Thus, the \textit{caloric} Appell transformation of the heat polynomials
(which, as seen, are associated with the eigenvalue problem for $\widehat{%
\mathrm{K}}_3$ with $\lambda =-i(n+1/2)$, $n=0,1,2,\dots$) yields the
temperature functions associated with the eigenvalue problem for the Laplace
dual operator $\widehat{\widetilde{\mathrm{K}}}_3=\widehat{\mathsf{T}}_{%
\mathbf{L}}\widehat{\mathrm{K}}_3\widehat{\mathsf{T}}_{\mathbf{L}}^{-1}$, or
equivalently with the eigenvalue problem for $\widehat{\mathrm{K}}_3$ with
opposite eigenvalues.

Notably, since the Fourier and Laplace dual operators of $\widehat{\mathrm{K}%
}_3$ are the same: $\widehat{\mathcal{F}}\widehat{\mathrm{K}}_3\widehat{%
\mathcal{F}}^{-1}=\widehat{\mathsf{T}}_{\mathbf{L}}\widehat{\mathrm{K}}_3%
\widehat{\mathsf{T}}_{\mathbf{L}}^{-1}=-$ $\widehat{\mathrm{K}}_3$, as
discussed in~\cite{torreat}, the Appell transformation of the heat
polynomials can as well be associated with a Fourier-similarity relation,
whilst in general, as seen above, one must resort to a Laplace-similarity
relation when dealing with generic temperature functions.

In order to complete the correspondence with the 2D PWE, we can say that a
\textit{fractional} caloric Appell transformation may also be introduced as
the evolving-time fractional Laplace transform operator
\[
\widehat{\mathcal{A}}^\alpha :=e^{-t\widehat{\mathrm{K}}_{-}}\widehat{%
\mathsf{T}}_{\mathbf{L}^\alpha }e^{t\widehat{\mathrm{K}}_{-}}=\widehat{%
\mathsf{T}}_{\mathbf{L}^\alpha }(t),
\]
with $\widehat{\mathsf{T}}_{\mathbf{L}^\alpha }$ given as discussed in Section~\ref{section3.1.1}. Evidently, it identif\/ies the transformation matrix
\[
\mathbf{M}_{\widehat{\mathcal{A}}^\alpha }=
\begin{pmatrix}
1 & -it \\
0 & 1
\end{pmatrix}
\begin{pmatrix}
\cos \phi & i\sin \phi \\
i\sin \phi & \cos \phi
\end{pmatrix}
\begin{pmatrix}
1 & it \\
0 & 1
\end{pmatrix} =
\begin{pmatrix}
\cos \phi +t\sin \phi & i(1+t^2)\sin \phi \\
i\sin \phi & \cos \phi -t\sin \phi
\end{pmatrix} ,
\]
and hence maps temperature functions into temperatures functions according
to
\[
w(x,t)=\widehat{\mathcal{A}}^\alpha v(x,t)=\frac 1{\sqrt{\cos \phi +t\sin
\phi }}e^{-\frac{\sin \phi }{2(\cos \phi +t\sin \phi )}x^2}v\left(\frac x{\cos
\phi +t\sin \phi },\frac{t\cos \phi -\sin \phi }{\cos \phi +t\sin \phi }\right).
\]

\subsection{Hankel-type transform and Appell transformation\\ for the radial
heat equation}\label{section4.4}

The radial HE is signif\/ied by
\begin{gather}
2\frac \partial {\partial t}u(r,t)=\widehat{D}_\mu u(r,t),  \label{rheat}
\end{gather}
where $\mu $ is an arbitrary real parameter and $\widehat{D}_\mu $ denotes
the dif\/ferential operator $\widehat{D}_\mu :=\frac{\partial ^2}{\partial r^2}%
+\frac{\mu -1}r\frac \partial {\partial r}$ \cite{bragg}. When $\mu =n$, a
positive integer, $\widehat{D}_n$ becomes the Laplacian operator in the
radial coordinate appropriate to a $n$-dimensional Euclidean space: $%
\widehat{D}_n=\frac{\partial ^2}{\partial x_1^2}+\cdots+\frac{\partial ^2}{%
\partial x_n^2}$. As before, we consider both $r\in [0,+\infty )$ and $t$ as
real unitless variables; also, as in (\ref{heat}), for computational
convenience we have added the factor $2$ with respect to the equation
appearing in the literature, thus implying the ``time'' scaling: $t\rightarrow t/2$ with respect to that equation.

For $\mu >1$, (\ref{rheat}) admits as solutions the radial heat polynomials $%
R_{n,\mu }(r,t)$, $n=0,1,2\dots$, and their Appell transforms $\widetilde{R}%
_{n,\mu }(r,t)$, explicitly given by \cite{bragg}
\begin{gather}
R_{n,\mu }(r,t):=2^nn!t^nL_n^{\mu /2-1}\left(-\frac{r^2}{2t}\right),\nonumber\\
\widetilde{R}_{n,\mu }(r,t):=S_\mu (r,t)R_{n,\mu }\left(\frac
rt,-\frac 1t\right)=t^{-2n}S_\mu (r,t)R_{n,\mu }(r,-t),\label{rhp}
\end{gather}
where $S_\mu (r,t)$ denotes the fundamental solution of (\ref{rheat}):
\begin{gather*}
S_\mu (r,t):=\frac 1{\left( 2\pi t\right) ^{\mu /2}}e^{-\frac{r^2}{2t}}.
\end{gather*}

With $\mu =2$ we can recognize a formal correspondence between (\ref{rheat})
and the 2D PWE (\ref{parrad}) with $m=0$, which so rules the evolution of
the radial wavefunction for a circularly symmetric wavef\/ield. Indeed, just
as solutions to the 2D radial PWE follow from the Hankel--Poisson integral (\ref{prad}) under def\/inite initial conditions, likewise solutions to~(\ref{rheat}) may be obtained from assigned initial conditions $u(r,0)=\varphi
(r) $ by the integral transform
\begin{gather}
u(r,t)=\int_0^{+\infty }\mathsf{K}_\mu (r,r^{\prime };t)\varphi (r^{\prime
})dr^{\prime },  \label{heattran}
\end{gather}
with
\begin{gather*}
\mathsf{K}_\mu (r,r^{\prime };t):=\frac 1tr^{1-\mu /2}r^{\prime  \mu
/2}e^{-\frac 1{2t}(r^2+r^{\prime  2})}I_{\mu /2-1}\left(\frac{rr^{\prime }}t\right),
%\label{kernelheat}
\end{gather*}
which yields, for instance, the heat polynomials $R_{n,\mu }(r,t)$ with $\varphi (r)=r^{2n}$.

More precisely, we may recognize a correspondence with the Hankel-type
transforms, discussed in Section~\ref{section3.3}, which, as there noted, for specif\/ic values
of the relevant parameters yield the conventional Hankel transform, or more
general (real and complex) radial transforms. In fact, on account of the
above clarif\/ied link between the 1D~HE and the Laplace transform, it is
natural to consider the kernel one would obtain from~(\ref{radial2}) with
the entries of the Laplace matrix~$\mathbf{L}$,
\begin{gather*}
\mathsf{K}_{_{\mathbf{L}}}^{(n,m)}(r,r^{\prime })=(-)^{m+n/2}(rr^{\prime
})^{1-n/2}I_{n/2+m-1}(rr^{\prime }),  %\label{radiallt}
\end{gather*}
which then, through (\ref{radial1}), should yield a sort of ``radial''
Laplace transform. More in general, using it in (\ref{radial1}) with the
aforestated correspondences $(n,m)\leftrightarrow (\nu ,\nu ^{\prime })$,
one would obtain in a~sense the complex versions of the transforms $\widehat{%
\mathcal{H}}_{_{1,\nu ,\nu ^{\prime }}}$ and $\widehat{\mathcal{H}}%
_{_{2,\nu ,\nu ^{\prime }}}$, i.e.
\begin{gather}
\big[ \widehat{\mathcal{L}}_{_{1,\nu ,\nu ^{\prime
}}}\varphi \big] (r):=(-)^{1+\nu }r^{1+2\nu ^{\prime }}\int_0^{+\infty
}(rr^{\prime })^{-\nu ^{\prime }}I_\nu (rr^{\prime })\varphi (r^{\prime
})dr^{\prime },\nonumber\\
\big[ \widehat{\mathcal{L}}_{_{2,\nu ,\nu ^{\prime }}}\varphi
\big] (r):=(-)^{1+\nu }\int_0^{+\infty }r^{\prime 1+2\nu ^{\prime
}}(rr^{\prime })^{-\nu ^{\prime }}I_\nu (rr^{\prime })\varphi (r^{\prime
})dr^{\prime }.\label{defhlt}
\end{gather}

Clearly, they relate to the Laplace transform just as the Barut--Girardello
transform (\ref{bgt}) relates to the Bargmann transform (\ref{bar}). In this
connection, we mention that in \cite{torreh}, reproducing the structure of
the transforms (\ref{defh}), two analogous Barut--Girardello-type transforms
have been introduced along with the relevant fractional-order versions.

It is easy to see that relation (\ref{expop}) yields for $\widehat{\mathcal{L%
}}_{_{1,\nu ,\nu ^{\prime }}}$ the operator representation
\begin{gather}
\widehat{\mathcal{L}}_{_{1,\nu ,\nu ^{\prime }}}=e^{-\widehat{\mathrm{K}}%
_{+}^{(1)}}e^{\widehat{\mathrm{K}}_{-}^{(1)}}e^{-\widehat{\mathrm{K}}%
_{+}^{(1)}}=e^{(\pi /2)[\widehat{\mathrm{K}}_{-}^{(1)}-\widehat{\mathrm{K}}%
_{+}^{(1)}]},  \label{rlexp}
\end{gather}
in terms of the operators (\ref{su1}); of course, a similar representation
can be obtained for the adjoint transform $\widehat{\mathcal{L}}_{_{2,\nu
,\nu ^{\prime }}}$ in terms of the operators (\ref{su2}).

The correspondence between (\ref{rlexp}) and (\ref{exptl}) is evident.

It is also evident that relation (\ref{expop}) (as that involving the
adjoint operator $\widehat{B}_{_{\nu ,\nu ^{\prime }}}$) can be made to
reproduce the general expression (\ref{heattran}) for the solutions of the
radial HE by setting $\nu =\mu /2-1$ and $\nu ^{\prime }=-\mu /2$ (as
correspondingly, $\nu =\nu ^{\prime }=\mu /2-1$). In fact, writing, for
instance:
\begin{gather}
\widehat{\mathcal{U}}_{_{\mathsf{RHE}}}(t):=e^{-t\widehat{\mathrm{K}}_{-}^{(1)}},  \label{heatdisp}
\end{gather}
with $\widehat{\mathrm{K}}_{-}^{(1)}=-\frac 12\widehat{B}_{_{\mu /2-1,-\mu
/2}}^{\dagger }=-\frac 12\widehat{D}_\mu $, we see that, by
\begin{gather*}
u(r,t)=\big[\widehat{\mathcal{U}}_{_{\mathsf{RHE}}}(t)\varphi \big](r),
%\label{heatdisp1}
\end{gather*}
one recovers (\ref{heattran}) on account of (\ref{expop}).

Then, just reproducing the considerations previously developed in connection
with the 2D PWE and the 1D~HE, in the light of the correspondence of (\ref{rlexp}) to (\ref{exptl}) and of (\ref{heatdisp}) to (\ref{eoheat}), we can
argue that the Appell transformation of solutions of the radial HE arises
from the action of the evolving-time ``radial-Laplace''-type transform
operator $\widehat{\mathcal{L}}_{_{1,\mu /2-1,-\mu /2}}(t)$ on temperature
functions. Explicitly, the relevant Appell operator $\widehat{\mathcal{A}}%
_\mu $ is to be intended as
\begin{gather}
\widehat{\mathcal{A}}_\mu :=e^{-t\widehat{\mathrm{K}}_{-}^{(1)}}\widehat{%
\mathcal{L}}_{_{1,\mu /2-1,-\mu /2}}e^{t\widehat{\mathrm{K}}_{-}^{(1)}}=%
\widehat{\mathcal{L}}_{_{1,\mu /2-1,-\mu /2}}(t),  \label{appellrheat}
\end{gather}
which amounts to the parallel ``heat conduction'' problems: $u(r,t)=e^{-t%
\widehat{\mathrm{K}}_{-}^{(1)}}\varphi (r)$ and $w(r,t)=e^{-t\widehat{%
\mathrm{K}}_{-}^{(1)}}\psi (r)$, with $\psi (r)=\widehat{\mathcal{L}}%
_{_{1,\mu /2-1,-\mu /2}}\varphi (r)$.

The direct evaluation or the use of the matrix-based procedure leads to the
well-known recipe (see (\ref{rhp})):
\begin{gather*}
\widehat{\mathcal{A}}_\mu u(r,t)=t^{-\mu /2}e^{-\frac{r^2}{2t}}u\left(\frac
rt,-\frac 1t\right)\propto S_\mu (r,t)u\left(\frac rt,-\frac 1t\right),  %\label{appellrheat1}
\end{gather*}
thus conf\/irming the interpretation (\ref{appellrheat}).

Needless to say, as before a fractional Appell transformation can as well be
introduced, by exploiting the fractional versions of the transforms~(\ref{defhlt}), inspired by those of the Hankel-type transforms~\cite{torreh}.

\section{Concluding notes}\label{section5}

After reviewing the interpretation of the \textit{optical} Appell
transformation \cite{torreat} as conveying the correspondence between
solutions of the 2D PWE, which evolve from Fourier or Hankel-related source
functions (according to whether the propagation problem concerns a
rectangular or a circular cylindrical geometry), a similar interpretation
has been shown to hold for the \textit{caloric} Appell transformation. It
manifests indeed the correspondence between temperature functions arising
from ``source functions'', which are related by a Laplace or a
``radial-Laplace''-type transform according to whether the 1D HE or the
radial HE is concerned. The \textit{optical} Appell transformation is a
symmetry transformation for the 2D PWE, which can be understood as an
evolving-location Fourier or Hankel transform operator. Similarly, the
\textit{caloric} Appell transformation is a symmetry transformation for the
HE, which can be regarded as an evolving-time Laplace or
``radial-Laplace''-type transform operator.

Also, resorting to the fractional versions of both the Fourier/Hankel and
Laplace/Hankel-type transforms, one can introduce a fractional Appell
transformation in relation to the PWE~\cite{torreat} as well as to the HE,
thus displaying a family of symmetry transformations for both equations,
parameterized by a continuous parameter.

The analysis has resorted to the Lie-algebra based method, as originally
developed in a series of seminal papers by Kalnins, Miller and Boyer~\cite{kalnins1,kalnins2,miller} in connection with the time-depen\-dent~SE. Also, it
echoes the analysis mastered in~\cite{torrel}, aimed at characterizing
transformations between wavefunctions in terms of transformations of the
respective source functions.

As earlier noted, the Appell transformation has relevance in other contexts
as well. It would be interesting to establish whether it is amenable for an
analogous interpretation even when framed in other contexts, like, for
instance, that of the Kolmogorov equation, considered in~\cite{brzezina}.

We conclude saying that other results concerned with the HE can be revisited
within the optical context. For instance, the property proven in \cite{leutwiler},
according to which the Appell transformation is \textit{essentially} the only transformation mapping solutions of the
(in general, $n $-dimensional) HE into solutions, applies to the optical context as well.
Thus, every transformation mapping wavefunctions into wavefunctions can be
understood as composed of Appell transformations, scalings and shifts of all the variables.
 Leutwiler's result has been restated in~\cite{torren} by
re\-sorting to the ray-matrix formalism of paraxial optics, and accordingly
exploiting suitable matrix factorization tools, like, for instance, the
possibility of realizing any $2\times 2$ optical ray-matrix, which
ultimately identif\/ies a symmetry operator for the 2D PWE, by an appropriate
sequence of free-sections and lenses, that then can be rearranged to yield
the Appell transformation matrix~(\ref{mappell}).

Since scalings and shifts of the inherent variables do not change the
functional form of the functions acted on by them, we can say that the
\textit{world} of the solutions of the 2D PWE can be understood as formed by
two classes of functions, which are just mapped one onto the other by the
Appell transform. Further solutions can be reached by scalings and/or shifts
of both the variables~$\xi $ and~$\zeta $, without changing evidently the
functional form of the original wavefunction.

{\sloppy This is not surprising, since, as noted in \cite{torren}, the symmetry
algebra pertaining to the PWE~\eqref{normpar} arises from Fourier-similarity related
operators, being in fact $\widehat{\mathcal{F}}\widehat{\mathrm{X}}\widehat{\mathcal{F}}^{-1}=-\widehat{\mathrm{P}}$, $\widehat{\mathcal{F}}\widehat{\mathrm{P}}\widehat{\mathcal{F}}^{-1}=\widehat{\mathrm{X}}$, and accordingly
$\widehat{\mathcal{F}}\widehat{\mathrm{K}}_{\pm }\widehat{\mathcal{F}}^{-1}=
\widehat{\mathrm{K}}_{\mp }$, $\widehat{\mathcal{F}}\widehat{\mathrm{K}}_3
\widehat{\mathcal{F}}^{-1}=-\widehat{\mathrm{K}}_3$.

}

Ultimately, the Appell transformation is a manifestation of such a property.

\subsection*{Acknowledgements}

The author wishes to thank an anonymous referee, who with his/her
suggestions, comments and criticisms has greatly improved the paper.

\pdfbookmark[1]{References}{ref}
\LastPageEnding


\begin{thebibliography}{99}

\footnotesize\itemsep=0pt

\bibitem{appell}
Appell M.P.,
Sur l'\'{e}quation $\frac{\partial ^2z}{\partial x^2}-\frac{\partial z}{\partial y}=0$ et la th\'{e}orie de la chaleur,
\textit{J. Math. Pure Appl.} \textbf{8} (1892), 187--216.

\bibitem{widderb}
Widder D.V.,
 The heat equation,
{\it Pure and Applied Mathematics}, Vol.~67, Academic Press, New York~-- London, 1975.

\bibitem{rosenbloom}
Rosenbloom P.C., Widder D.V.,
Expansions in terms of heat polynomials and associated functions,
\href{http://dx.doi.org/10.2307/1993155}{\textit{Trans. Amer. Math. Soc.}} \textbf{92} (1959), 220--266.

\bibitem{widder1}
Widder D.V.,
Analytic solutions of the heat equation,
\href{http://dx.doi.org/10.1215/S0012-7094-62-02950-2}{\textit{Duke Math.~J.}} \textbf{29} (1962), 497--503.

\bibitem{widder2}
Widder D.V.,
Expansions in series of homogeneous temperature functions of the f\/irst and second kinds,
\href{http://dx.doi.org/10.1215/S0012-7094-69-03660-6}{\textit{Duke Math.~J.}} \textbf{36} (1969), 495--509.

\bibitem{leutwiler}
Leutwiler H.,
On the Appell transformation, in Potential Theory (Prague, 1987),
Editors J.~Kr\`{a}l et. al., Plenum Press, New York, 1988, 215--222.

\bibitem{shimomura}
Shimomura  K.,
The determination of caloric morphisms on Euclidean domains,
\textit{Nagoya Math.~J.} \textbf{158} (2000), 133--166.

\bibitem{brzezina}
Brzezina M.,
Appell type transformation for the Kolmogorov operator,
\href{http://dx.doi.org/10.1002/mana.19941690105}{\textit{Math. Nachr.}} \textbf{169} (1994), 59--67.

\bibitem{torreat}
Torre A.,
The Appell transformation for the paraxial wave equation,
\href{http://dx.doi.org/10.1088/2040-8978/13/1/015701}{\textit{J.~Opt.}} \textbf{13} (2011), 015701, 14~pages.

\bibitem{kalnins1}
Kalnins E.G.,  Miller W.~Jr.,
 Lie theory and separation of variables. V.~The equation $iU_t+U_{xx}=0$ and $iU_t+U_{xx}-(c/x^2)U=0$,
\href{http://dx.doi.org/10.1063/1.1666533}{\textit{J.~Math. Phys.}} \textbf{15} (1974), 1728--1737.

\bibitem{kalnins2}
Boyer C.P., Kalnins E.G., Miller  W. Jr.,
Lie theory and separation of variables. VI.~The equation $iU_t+\Delta _2U=0$,
\href{http://dx.doi.org/10.1063/1.522573}{\textit{J.~Math. Phys.}} \textbf{16} (1975), 499--511.

\bibitem{miller}
Miller W.~Jr.,
Symmetry and separation of variables,
{\it Encyclopedia of Mathematics and its Applications}, Vol.~4, Addison-Wesley Publishing Co., Reading, Mass.~-- London~-- Amsterdam, 1977.

\bibitem{wolfb}
Wolf K.B.,
Integral transforms in science and engineering,
{\it Mathematical Concepts and Methods in Science and Engineering}, Vol.~11, Plenum Press, New York~-- London, 1979.

\bibitem{olver}
Olver P.J.,
Applications of Lie groups to dif\/ferential equations,
2nd ed., {\it Graduate Texts in Mathematics}, Vol.~107, Springer-Verlag, New York, 1993.

\bibitem{torrepwe}
Torre A.,
A note on the general solution of the paraxial wave equation: a~Lie algebra view,
\href{http://dx.doi.org/10.1088/1464-4258/10/5/055006}{\textit{J.~Opt.~A: Pure Appl. Opt.}} \textbf{10} (2008), 055006, 14~pages.

\bibitem{torresp}
Torre A.,
Separable-variable solutions of the wave equation from a general type of solutions of the paraxial wave equation,
in Proceedings of the International Conference ``Days on Dif\/fraction'' (May 26--29, 2009, St. Petersburg), 178--183.

\bibitem{torrel}
Torre A.,
Linear and quadratic exponential modulation of the solutions of the paraxial wave equation,
\href{http://dx.doi.org/10.1088/2040-8978/12/3/035701}{\textit{J.~Opt.}} \textbf{12} (2010), 035701, 11~pages.

\bibitem{torren}
 Torre A.,
 Appell transformation and symmetry transformations for the paraxial wave equation,
\href{http://dx.doi.org/10.1088/2040-8978/13/7/075710}{\textit{J.~Opt.}} \textbf{13} (2011), 075710, 12~pages.

\bibitem{kato}
Kato T.,
Perturbation theory for linear operators,
2nd ed., {\it Grundlehren der Mathematischen Wissenschaften}, Band~132, Springer-Verlag, Berlin~-- New York, 1976.

\bibitem{collins}
Collins S.A.~Jr.,
Lens-system dif\/fraction integral written in terms of matrix optics,
\href{http://dx.doi.org/10.1364/JOSA.60.001168}{\textit{J.~Opt. Soc. Amer.~A}} \textbf{60} (1970), 1168--1177.

\bibitem{siegman}
Siegman A.E.,
Lasers, University Science Books, Mill Valley, CA, 1986.


\bibitem{ballantine}
Ballentine L.E.,
Quantum mechanics,
Prentice Hall, Englewood Clif\/fs, New Jersey, 1990.

\bibitem{bandresc}
Bandres M.A., Guti\'{e}rrez-Vega J.C.,
Cartesian beams,
\href{http://dx.doi.org/10.1364/OL.32.003459}{\textit{Opt. Lett.}} \textbf{32} (2007), 3459--3461.

\bibitem{bandresr}
Bandres M.A., Guti\'{e}rrez-Vega J.C.,
Circular beams,
\href{http://dx.doi.org/10.1364/OL.33.000177}{\textit{Opt. Lett.}} \textbf{33} (2008), 177--179.

\bibitem{bandrese}
Bandres M.A., Guti\'{e}rrez-Vega J.C.,
Elliptical beams,
\href{http://dx.doi.org/10.1364/OE.16.021087}{\textit{Opt. Expr.}} \textbf{16} (2008), 21087--21092.

\bibitem{sudarshan}
Sudarshan E.C.G., Mukunda N., Simon R.,
Realization of f\/irst order optical systems using thin lenses,
\textit{Opt. Acta} \textbf{32} (1985), 855--872.

\bibitem{bandrespg}
Bandres M.A., Guizar-Sicairos M.,
Paraxial group,
\href{http://dx.doi.org/10.1364/OL.34.000013}{\textit{Opt. Lett.}} \textbf{34} (2009), 13--15.

\bibitem{wei}
Wei J., Norman E.,
Lie algebraic solution of linear dif\/ferential equations,
\href{http://dx.doi.org/10.1063/1.1703993}{\textit{J.~Math. Phys.}} \textbf{4} (1963), 575--581.\\
 Dattoli G., Gallardo J.C., Torre A.,
 An algebraic view to the operatorial ordering and its applications to optics,
 \textit{Riv. Nuovo Cimento~(3)} \textbf{11} (1988), no.~11, 1--79.\\
  Ban M.,
  Decomposition formulas for $su(1,1)$ and $su(2)$ Lie algebras and their applications to quantum optics,
\href{http://dx.doi.org/10.1364/JOSAB.10.001347}{\textit{J.~Opt. Soc. Amer.~B}} \textbf{10} (1993), 1347--1359.

\bibitem{magnus}
Magnus W., Oberhettinger F., Soni R.P.,
Formulas and theorems for the special functions of mathematical physics,
3rd ed., {\it Die Grundlehren der mathematischen Wissenschaften}, Band~52, Springer-Verlag, New York, 1966.

\bibitem{torrea}
Torre A.,
A note on the Airy beam in the light of the symmetry algebra based approach,
\href{http://dx.doi.org/10.1088/1464-4258/11/12/125701}{\textit{J.~Opt.~A: Pure Appl. Opt.}} \textbf{11} (2009), 125701, 11~pages.

\bibitem{berry}
Berry M.V., Balazs N.L.,
Nonspreading wave packets,
\href{http://dx.doi.org/10.1119/1.11855}{\textit{Amer.~J. Phys.}} \textbf{47} (1979), 264--267.

\bibitem{besieris}
Besieris I.M., Shaarawi A.M., Ziolkowski R.W.,
Nondispersive accelerating wave packets,
\href{http://dx.doi.org/10.1119/1.17510}{\textit{Amer.~J. Phys.}} \textbf{62} (1994), 519-521.

\bibitem{bandresa}
Bandres M.A.,
Accelerating beams,
\href{http://dx.doi.org/10.1364/OL.34.003791}{\textit{Opt. Lett.}} \textbf{34} (2009), 3791--3793.

\bibitem{siviloglou}
Siviloglou G.A., Christodoulides D.N.,
Accelerating f\/inite energy Airy beams,
\href{http://dx.doi.org/10.1364/OL.32.000979}{\textit{Opt. Lett.}} \textbf{32} (2007), 979--981.\\
 Siviloglou G.A., Broky J., Dogariu A.,  Christodoulides D.N.,
 Observation of accelerating Airy beams,
\href{http://dx.doi.org/10.1103/PhysRevLett.99.213901}{\textit{Phys. Rev. Lett.}} \textbf{99} (2007), 2139011, 4~pages.\\
 Siviloglou G.A., Broky J., Dogariu A., Christodoulides D.N.,
 Ballistic dynamics of Airy beams,
 \href{http://dx.doi.org/10.1364/OL.33.000207}{\textit{Opt. Lett.}} \textbf{33} (2008), 207--209.

\bibitem{besieris1}
Besieris I.M., Shaarawi A.M.,
A note on an accelerating f\/inite energy Airy beam,
\href{http://dx.doi.org/10.1364/OL.32.002447}{\textit{Opt. Lett.}} \textbf{32} (2007), 2447--2449.

\bibitem{broky}
Broky J., Siviloglou G.A., Dogariu A., Christodoulides D.N.,
Self-healing properties of optical Airy beams,
\href{http://dx.doi.org/10.1364/OE.16.012880}{\textit{Opt. Expr.}} \textbf{16} (2008), 12880--12891.

\bibitem{morris}
Morris J.E., Mazilu M., Baumgartl J., Cizmar T., Dholakia K.,
Propagation characteristics of Airy beams: dependence upon spatial coherence and wavelength,
\href{http://dx.doi.org/10.1364/OE.17.013236}{\textit{Opt. Expr.}} \textbf{17} (2009), 13236--13245.

\bibitem{dai}
Dai H.T., Sun X.W., Luo D., Liu Y.J.,
Airy beams generated by binary phase element made of polymer-dispersed liquid crystals,
\href{http://dx.doi.org/10.1364/OE.17.019365}{\textit{Opt. Expr.}} \textbf{17} (2009), 19365--19370.

\bibitem{baumgartl}
Baumgartl J., Mazilu  M.,  Dholakia K.,
Optically mediated particle clearing using Airy wavepackets,
\href{http://dx.doi.org/10.1038/nphoton.2008.201}{\textit{Nature Photonics}} \textbf{2} (2008), 675--678.

\bibitem{ellenbogen}
Ellenbogen T., Voloch-Bloch N., Ganany-Padowicz A., Arie A.,
Nonlinear generation and manipulation of Airy beams,
\href{http://dx.doi.org/10.1038/nphoton.2009.95}{\textit{Nature Photonics}} \textbf{3} (2009), 395--398.

\bibitem{salandrino}
Salandrino A., Christodoulides D.N.,
Airy plasmon: a nondif\/fracting surface wave,
\href{http://dx.doi.org/10.1364/OL.35.002082}{\textit{Opt. Lett.}} \textbf{35} (2010), 2082--2084.

\bibitem{frt}
Mendlovic D., Ozaktas H.M.,
Fractional Fourier transforms and their optical implementation.~I,
\href{http://dx.doi.org/10.1364/JOSAA.10.001875}{\textit{J.~Opt. Soc. Amer.~A}}  \textbf{10} (1993), 1875--1881.\\
 Mendlovic D.,  Ozaktas H.M.,
Fractional Fourier transforms and their optical implementation.~II,
\href{http://dx.doi.org/10.1364/JOSAA.10.002522}{\textit{J.~Opt. Soc. Amer.~A}} \textbf{10} (1993), 2522--2531.\\
 Ozaktas H.M., Kutay  M.A., Mendlovic D.,
 Introduction to the fractional Fourier transform and its applications,
in \href{http://dx.doi.org/10.1016/S1076-5670(08)70272-6}{{\it Advances in Imaging and Electron Physics}}, Vol.~106, Editor P.W.~Hawkes, Academic Press, San Diego, 1999, 239--291.

\bibitem{ozaktas}
Ozatkas H.M., Zalevsky  Z., Kutay M.A.,
The fractional Fourier transform with applications in optics and signal processing, Wiley, New York, 2001.

\bibitem{lohmann}
Lohmann A.W.,
Image rotation, Wigner rotation, and the fractional order Fourier transform,
\href{http://dx.doi.org/10.1364/JOSAA.10.002181}{\textit{J. Opt. Soc. Amer.~A}}  \textbf{10} (1993), 2181--2186.\\
 Lohmann A.W., Mendlovic D., Zalevsky Z.,
Fractional transformations in optics,
in \href{http://dx.doi.org/10.1016/S0079-6638(08)70352-4}{\textit{Progress in Optics}}, Vol.~38, Editor E.~Wolf, Elsevier, Amsterdam, 1997, 263--342.

\bibitem{torrepo}
Torre A.,
The fractional Fourier transform and some of its applications to optics,
 in \href{http://dx.doi.org/10.1016/S0079-6638(02)80031-2}{\textit{Progress in Optics}}, Vol.~43, Editor E.~Wolf, Elsevier, Amsterdam, 2002,  531--596.

\bibitem{durnin}
Durnin J.,
Exact solutions for nondif\/fracting beams. I.~The scalar theory,
\href{http://dx.doi.org/10.1364/JOSAA.4.000651}{\textit{J. Opt. Soc. Amer.~A}} (1987), 651--654.\\
Durnin J., Miceli J.J.,  Eberly J.H., Dif\/fraction-free beams,
\href{http://dx.doi.org/10.1103/PhysRevLett.58.1499}{\textit{Phys. Rev. Lett.}} \textbf{58} (1987), 1499--1501.

\bibitem{sheppardbg}
Sheppard C.J.R.,  Wilson T.,
Gaussian beams theory of lenses with annular aperture,
\textit{IEE J. Microwaves Opt. Acoust.} \textbf{2} (1978), 105--112,\\
 Gori F., Guattari G., Padovani C.,
 Bessel--Gauss beams,
\href{http://dx.doi.org/10.1016/0030-4018(87)90276-8}{\textit{Opt. Comm.}} \textbf{64} (1987), 491--495.

\bibitem{lohmannd}
Lohmann A.W.,
Ein neues Dualitatsprinzip in der Optik,
\textit{Optik} \textbf{11} (1954), 478--488.\\
 Lohmann A.W.,
 Duality in optics, \textit{Optik} \textbf{89} (1992), 93--97.

\bibitem{sheppard}
Sheppard C.J.R.,
Beam duality, with application to generalized Bessel--Gaussian, and Hermite-- and Laguerre--Gaussian beams,
\href{http://dx.doi.org/10.1364/OE.17.003690}{\textit{Opt. Expr.}} \textbf{15} (2009), 3690--3697.

\bibitem{heaviside1}
Heaviside O.,
Electrical papers, The Macmillan Co., New York and London, 1892.\\
 Heaviside O., Electromagnetic theory, The Electrician Printing \& Publishing Co., London, Vol.~1, 1894; Vol.~2, 1899;
Vol.~3, 1912.

\bibitem{heaviside2}
Nahin P.J.,
Oliver Heaviside: the life, work, and times of an electrical genius of the victorian age,
The Johns Hopkins University Press, Baltimore, 2002.

\bibitem{tranter}
Tranter C.J.,
Integral transforms in mathematical physics,
Methuen \& Co., Ltd., London; John Wiley \& Sons, Inc., New York, 1951.

\bibitem{davies}
Davies B.,
Integral transforms and their applications,
{\it Applied Mathematical Sciences}, Vol.~25, Springer-Verlag, New York~-- Heidelberg, 1978.

\bibitem{bargmann}
Bargmann V.,
On a Hilbert space of analytic functions and an associated integral transform,
\href{http://dx.doi.org/10.1002/cpa.3160140303}{\textit{Comm. Pure Appl. Math.}} \textbf{14} (1961), 187--214.\\
 Bargmann V.,
 On a Hilbert space of analytic functions and an associated integral transform. II.~A~family of related function spaces. Application to distribution theory,
\href{http://dx.doi.org/10.1002/cpa.3160200102}{\textit{Comm. Pure Appl. Math.}} \textbf{20} (1967), 1--101.

\bibitem{moshinsky}
Moshinsky M., Quesne C.,
Linear canonical transformations and their unitary representations,
\href{http://dx.doi.org/10.1063/1.1665805}{\textit{J.~Math. Phys.}} \textbf{12} (1971), 1772--1780.

\bibitem{quesne}
Quesne C., Moshinsky M.,
Canonical transformations and matrix elements,
\href{http://dx.doi.org/10.1063/1.1665806}{\textit{J.~Math. Phys.}} \textbf{12} (1971), 1780--1783.

\bibitem{wolfl}
Wolf K.B.,
Canonical transforms. I.~Complex linear transforms,
\href{http://dx.doi.org/10.1063/1.1666811}{\textit{J.~Math. Phys.}} \textbf{15} (1974), 1295--1301.

\bibitem{wolfr}
Wolf K.B.,
Canonical transforms. II.~Complex radial transforms,
\href{http://dx.doi.org/10.1063/1.1666590}{\textit{J.~Math. Phys.}} \textbf{15} (1974), 2102--2111.

\bibitem{kramer}
Kramer P., Moshinsky M., Seligman T.H.,
Complex extensions of canonical transformations and quantum mechanics,
in Group Theory and Its Applications, Vol.~III, Editor E.M.~Loebl, Academic Press, New York, 1975, 249--332.

\bibitem{wolfpde}
Wolf K.B.,
Canonical transforms, separation of variables and similarity solutions for a class of parabolic dif\/ferential equations,
\href{http://dx.doi.org/10.1063/1.522951}{\textit{J.~Math. Phys.}} \textbf{17} (1976), 601--613.

\bibitem{wolfef}
Wolf K.B.,
On self-reciprocal functions under a class of integral transforms,
\href{http://dx.doi.org/10.1063/1.523365}{\textit{J.~Math. Phys.}} \textbf{18} (1977), 1046--1051.

\bibitem{namias1}
Namias V.,
The fractional order Fourier transform and its application to quantum mechanics,
\href{http://dx.doi.org/10.1093/imamat/25.3.241}{\textit{J.~Inst. Appl. Math.}} \textbf{25} (1980), 241--265.

\bibitem{namias2}
Namias V.,
Fractionalization of Hankel transforms,
\href{http://dx.doi.org/10.1093/imamat/26.2.187}{\textit{J.~Inst. Math. Appl.}} \textbf{26} (1980), 187--197.

\bibitem{mcbride}
McBride A.C., Kerr F.H.,
On Namias's fractional Fourier transforms,
\href{http://dx.doi.org/10.1093/imamat/39.2.159}{\textit{IMA J. Appl. Math.}} \textbf{39} (1987), 159--175.

\bibitem{pei}
Pei S.-C., Ding J.-J.,
Eigenfunctions of linear canonical transform,
\href{http://dx.doi.org/10.1109/78.972478}{\textit{IEEE Trans. Signal Process.}} \textbf{50} (2002), 11--26.

\bibitem{torretr}
Torre~A.,
Linear and radial transforms of fractional order,
\href{http://dx.doi.org/10.1016/S0377-0427(02)00637-4}{\textit{J.~Comp. Appl. Math.}} \textbf{153} (2003), 477--486.

\bibitem{alieva}
Alieva T., Bastiaans M.J.,
Properties of the linear canonical integral transformation,
\href{http://dx.doi.org/10.1364/JOSAA.24.003658}{\textit{J.~Opt. Soc. Amer.~A}} \textbf{24} (2007), 3658--3665.

\bibitem{stern}
Stern A.,
Uncertainty principles in linear canonical transform domains and some of their implications in optics,
\href{http://dx.doi.org/10.1364/JOSAA.25.000647}{\textit{J.~Opt. Soc. Amer.~A}} \textbf{25} (2008), 647--652.

\bibitem{deng}
Deng B., Tao R., Wang Y.,
Convolution theorems for the linear canonical transform and their applications,
\href{http://dx.doi.org/10.1007/s11432-006-2016-4}{\textit{Sci. China Ser.~F}} \textbf{49} (2006), 592--603.

\bibitem{koc}
Ko\c{c} A., Ozaktas H.M., Hesselink L.,
Fast and accurate algorithm for the computation of complex linear canonical transforms,
\href{http://dx.doi.org/10.1364/JOSAA.27.001288}{\textit{J.~Opt. Soc. Amer.~A}}  \textbf{27} (2010), 1288--1302.

\bibitem{sharma}
Sharma K.K.,
Fractional Laplace transform,
\href{http://dx.doi.org/10.1007/s11760-009-0127-2}{\textit{Signal Image Video Process}} \textbf{4} (2010), 377--379.

\bibitem{louck}
Louck J.D., Moshinsky M., Wolf K.B.,
Canonical transformations and accidental degeneracy. I.~The anisotropic oscillator,
\href{http://dx.doi.org/10.1063/1.1666379}{\textit{J.~Math. Phys.}} \textbf{14} (1973), 692--695,\\
 Louck J.D., Moshinsky M., Wolf K.B.,
 Canonical transformations and accidental degeneracy. II.~The isotropic oscillator in a sector,
\href{http://dx.doi.org/10.1063/1.1666380}{\textit{J.~Math. Phys.}} \textbf{14} (1973), 696--700.

\bibitem{barut}
Barut A.O., Girardello L.,
New ``coherent'' states associated with non-compact groups,
\href{http://dx.doi.org/10.1007/BF01646483}{\textit{Comm. Math. Phys.}} \textbf{21} (1971), 41--55.

\bibitem{linares}
Linares Linares M., M\'endez P\'erez J.M.R.,
A~Hankel type integral transformation on certain space of distributions,
\textit{Bull. Calcutta Math. Soc.} \textbf{83} (1991), 447--546.\\
Linares Linares  M., M\'endez P\'erez J.M.R.,
Hankel complementary integral transformations of arbitrary order,
\href{http://dx.doi.org/10.1155/S0161171292000401}{\textit{Internat.~J. Math. Math. Sci.}} \textbf{15} (1992), 323--332.

\bibitem{malgonde2}
Malgonde S.P.,  Debnath L.,
On Hankel type integral transformations of generalized functions,
\href{http://dx.doi.org/10.1080/10652460410001686055}{\textit{Integral Transforms Spec. Funct.}} \textbf{15} (2004), 421--430.

\bibitem{malgonde3}
Malgonde S.P., Bandewar S.R.,  Debnath L.,
Mixed Parseval equation and generalized Hankel-type integral transformation of distributions,
\href{http://dx.doi.org/10.1080/10652460410001686046}{\textit{Integral Transforms Spec. Funct.}} \textbf{15} (2004), 431--443.

\bibitem{torreh}
Torre A.,
Hankel-type integral transforms and their fractionalization: a note,
\href{http://dx.doi.org/10.1080/10652460701827848}{\textit{Integral Transforms Spec. Funct.}} \textbf{19} (2008), 277--292.

\bibitem{bragg}
Bragg L.R.,
The radial heat polynomials and related functions,
\href{http://dx.doi.org/10.2307/1994051}{\textit{Trans. Amer. Math. Soc.}} \textbf{119} (1965), 270--290.\\
 Bragg L.R.,
 The radial heat equation and Laplace transforms,
\href{http://dx.doi.org/10.1137/0114080}{\textit{SIAM~J. Appl. Math.}} \textbf{14} (1966), 986--993.

\end{thebibliography}
\end{document}